\RequirePackage{fix-cm}
\documentclass[twocolumn,epjc3]{svjour3}
\smartqed  
\RequirePackage{graphicx}
\journalname{Eur. Phys. J. C}

\begin{document}

\title{Analysis of backgrounds for the ANAIS-112 dark matter experiment}

\author{J.~Amar\'{e}\thanksref{addr1,addr2}
        \and S.~Cebri\'{a}n\thanksref{e1,addr1,addr2}
        \and I.~Coarasa\thanksref{addr1,addr2}
        \and C.~Cuesta\thanksref{addr1,addr4}
        \and E.~Garc\'{i}a\thanksref{addr1,addr2}
        \and M.~Mart\'{i}nez\thanksref{addr1,addr2,addr3}
        \and M.A.~Oliv\'{a}n\thanksref{addr1,addr5}
        \and Y.~Ortigoza\thanksref{addr1,addr2}
        \and A.~Ortiz de Sol\'{o}rzano\thanksref{addr1,addr2}
        \and J.~Puimed\'{o}n\thanksref{addr1,addr2}
        \and A.~Salinas\thanksref{addr1,addr2}
        \and M.L.~Sarsa\thanksref{addr1,addr2}
        \and J.A.~Villar\thanksref{e2,addr1,addr2}
        \and P.~Villar\thanksref{addr1,addr2}
        }

\thankstext{e1}{e-mail: scebrian@unizar.es}
\thankstext{e2}{Deceased}

\institute{Laboratorio de F\'{i}sica Nuclear y Astropart\'{i}culas, Universidad de Zaragoza, Calle Pedro Cerbuna 12, 50009 Zaragoza, Spain \label{addr1}
           \and
           Laboratorio Subterr\'{a}neo de Canfranc, Paseo de los Ayerbe s/n, 22880 Canfranc Estaci\'{o}n, Huesca, Spain\label{addr2}
           \and
           Fundaci\'{o}n Agencia Aragonesa para la Investigaci\'{o}n y el Desarrollo, ARAID, Gobierno de Arag\'{o}n, Avenida de Ranillas 1-D, 50018 Zaragoza, Spain\label{addr3}
           \and
           \emph{Present Address:} Centro de Investigaciones Energ\'{e}ticas, Medioambientales y Tecnol\'{o}gicas, CIEMAT, 28040, Madrid, Spain\label{addr4}
           \and
           \emph{Present Address:} Fundaci\'{o}n CIRCE, 50018, Zaragoza, Spain\label{addr5}
}
\date{Received: date / Accepted: date}
\maketitle

\begin{abstract}
The ANAIS (Annual modulation with NaI(Tl) Scintillators) experiment aims at the confirmation or refutation of the DAMA/LIBRA positive annual modulation signal in the low energy detection rate, using the same target and technique, at the Canfranc Underground Laboratory (LSC) in Spain. ANAIS-112, consisting of nine 12.5~kg NaI(Tl) modules produced by Alpha Spectra Inc. in a 3$\times$3 matrix configuration, is taking data smoothly in ``dark matter search'' mode since August, 2017, after a commissioning phase and operation of the first detectors during the last years in various set-ups. A large effort has been carried out within ANAIS to characterize the background of sodium iodide detectors, before unblinding the data and performing the first annual modulation analysis. Here, the background models developed for all the nine ANAIS-112 detectors are presented. Measured spectra from threshold to high energy in different conditions are well described by the models
based on quantified activities independently estimated following several approaches. In the region from 1 to 6~keV$_{ee}$ the measured, efficiency corrected background level is 3.58$\pm$0.02~keV$^{-1}$ kg$^{-1}$ d$^{-1}$; NaI crystal bulk contamination is the dominant background source being $^{210}$Pb, $^{40}$K, $^{22}$Na and $^{3}$H contributions the most relevant ones. This background level, added to the achieved 1~keV$_{ee}$ analysis threshold (thanks to the outstanding light collection and robust filtering procedures developed), allow ANAIS-112 to be sensitive to the modulation amplitude measured by DAMA/LIBRA, and able to explore at three sigma level in five years the WIMP parameter region singled out by this experiment.
\end{abstract}

\section{Introduction}
Experimental efforts looking for Weakly Interacting Massive Particles (WIMPs), one of the preferred categories of the hypothetical dark matter particles our galactic halo should consist of in order to explain the Milky Way's rotation curve, date back to the eighties of the last century; sensitivities have been continuously improving since then, profiting from the application of ultra-low background techniques in the building of the detectors, and the development of new detection schemes. The DAMA/LIBRA experiment, running first 100~kg (DAMA/NaI phase) and later 250~kg (DAMA/LIBRA phases 1 and 2) of NaI(Tl) detectors at the Gran Sasso Underground Laboratory, has found a significant (12.9$\sigma$) rate modulation over 20 years in the 2-6~keV$_{ee}$\footnote{Electron-equivalent energy. In the following, we will use just keV for keV$_{ee}$.} energy region showing the expected features for a galactic halo WIMPs signal \cite{bernabei13,bernabei18}. 
Other experiments searching for dark matter with other targets or techniques have not found any signal of dark matter \cite{cdmsresults,edelweissresults,cresstresults,xenonresults,luxresults,pandaxresults,darksideresults,picoresults,damicresults,newsgresults} and have been ruling out the most plausible compatibility dark matter scenarios \cite{kang}, although comparison among results obtained with different target nuclei is model dependent. The identification of the annual modulation in the detection rate is a signature of the scattering of the dark matter particles with the detector nuclei \cite{freese}; other experiments using different targets have not presented evidence of this effect \cite{xenonmodulation,xmassmodulation,luxmodulation}, although some hints were reported \cite{cogentmodulation}. A model-independent confirmation of the annual modulation positive signal reported by DAMA/LIBRA using the same target and technique, but different experimental conditions, is therefore of utmost importance at present. An independent experiment, having different residual cosmic ray flux and environmental conditions than DAMA/LIBRA has, could bring a new insight in the DAMA/LIBRA observation: confirmation of a modulation having the same phase and amplitude would be very difficult to explain as effect of backgrounds or systematics.

Although sodium iodide crystals doped with Tl have been applied in the direct search of galactic dark matter particles for a long time \cite{fushimi99,sarsa97,bernabei99,gerbier99,naiad}, only  recently radiopurity levels in NaI(Tl) crystals are close to those achieved by DAMA/LIBRA detectors, allowing to bring some light into a long-standing controversy. ANAIS (Annual modulation with NaI Scintillators) \cite{anais17} aims at the study of the dark matter annual modulation with ultrapure NaI(Tl) scintillators at the Canfranc Underground Laboratory (LSC) in Spain to test the DAMA/LIBRA result. ANAIS-112 uses a NaI(Tl) mass of 112.5~kg. ANAIS approach is also pursued by COSINE \cite{cosine,cosinenature}, merging KIMS \cite{kims} and DM-Ice \cite{dmice} collaborations, using crystals from the same ANAIS provider, and on a longer time basis by SABRE \cite{sabre} and COSINUS \cite{cosinus} collaborations, which are working on their own developments on radiopure NaI(Tl).

Robust background models are essential for experiments demanding ultra low background conditions, in order to guide the design, to analyze any possible systematics and to make reliable estimates of the experiment sensitivity (see some examples at \cite{xenon100bkg,edelweissbkg,gerdabkg,luxbkg,exobkg,cuorebkg,cosinebkg}). Following the first quantification of cosmogenic radionuclide production and its effects in NaI(Tl) crystals \cite{anaisjcap} and background assessment using data from the first ANAIS-112 modules produced by Alpha Spectra \cite{anaisbkg}, a complete analysis and quantification of the different background components in the whole ANAIS-112 experiment is presented here. It is worth noting that the reliability of this study is based on an accurate assay of background sources, a careful computation of their contribution to the experiment (made by Monte Carlo simulation) and continuous validation of the obtained results against experimental data, in different experimental conditions and energy ranges.

The structure of the article is the following. The experimental set-up of the ANAIS-112 experiment and the measurements used in this study are described in section~\ref{setup}. The quantification of the primordial and cosmogenic activity in the NaI(Tl) crystals and other components, following different approaches, is presented in section~\ref{quanti}. Section \ref{modeling}
shows the details of the background simulation. Then, the background models obtained and their validation against experimental data at different
conditions are discussed in section~\ref{comparison}. Finally, conclusions are summarized in
section~\ref{conclusions}.

\section{The ANAIS-112 experiment} \label{setup}

ANAIS-112 is the result of an extensive effort to achieve the best detector performance and background understanding using NaI(Tl) detectors from different providers. To confirm the DAMA/LIBRA result, ANAIS detectors should be comparable to those of DAMA/LIBRA in terms of energy threshold (1-2~keV) and radioactive background (1-2~counts/(keV~kg~d) in the region of interest (RoI) below 6~keV). Several prototypes using BICRON and Saint-Gobain crystals \cite{anaisap,anaisepjc} were developed and operated at LSC, but disregarded due to an unacceptable K content in the crystal. In the end, detectors built by Alpha Spectra Inc. were selected for final configuration because of the outstanding optical quality, although background goals could not be fully achieved. In the last years, the first modules of ANAIS-112 have been operated in various set-ups at the LSC \cite{anaisnima,anaispprocedia} in order to understand the background components and to characterize their response; an outstanding light collection at the level of 15~phe/keV has been measured for all detectors \cite{anaisyield,MAthesis,anaiscompanion}, which is essential to guarantee a low energy threshold.

ANAIS-112 consists of nine modules, 12.5~kg each, of ultrapure sodium iodide Thallium doped, NaI(Tl), built by the Alpha Spectra company in Colorado (US) along several years (see table~\ref{crystals}). All the crystals are cylindrical (4.75'' diameter and 11.75'' length) and are housed in OFE (Oxygen Free Electronic) copper. A Mylar window in the lateral face allows for low energy calibration. Crystals were coupled through synthetic quartz windows to two Hamamatsu photomultipliers (PMTs) at the LSC clean room in a second step. Different PMT models were tested in order to choose the best option in terms of light collection and background \cite{Clarathesis}, selecting finally the R12669SEL2 model.

\begin{figure*}
\centering
 \includegraphics[width=0.6\textwidth]{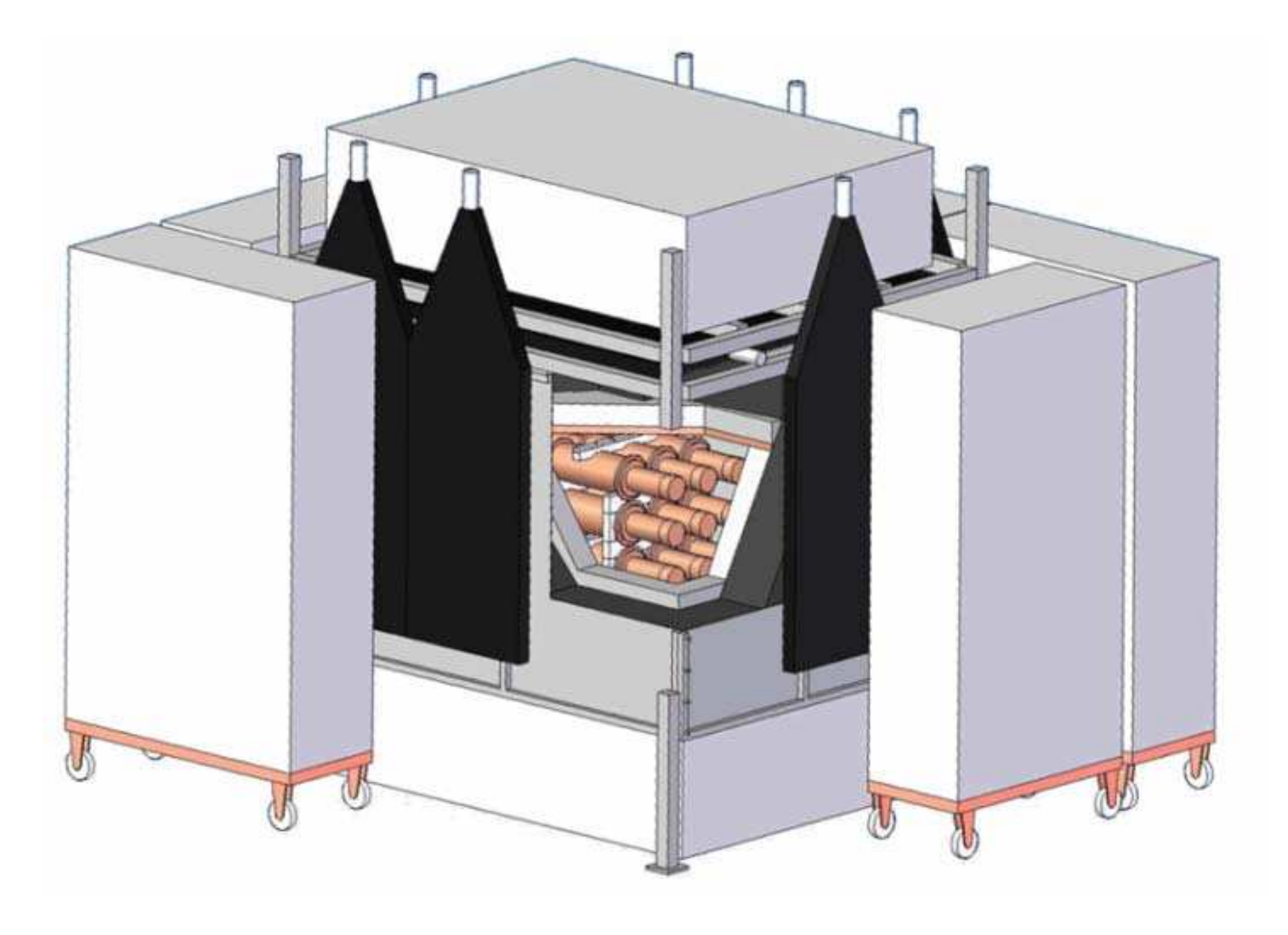}
 \includegraphics[width=0.3\textwidth]{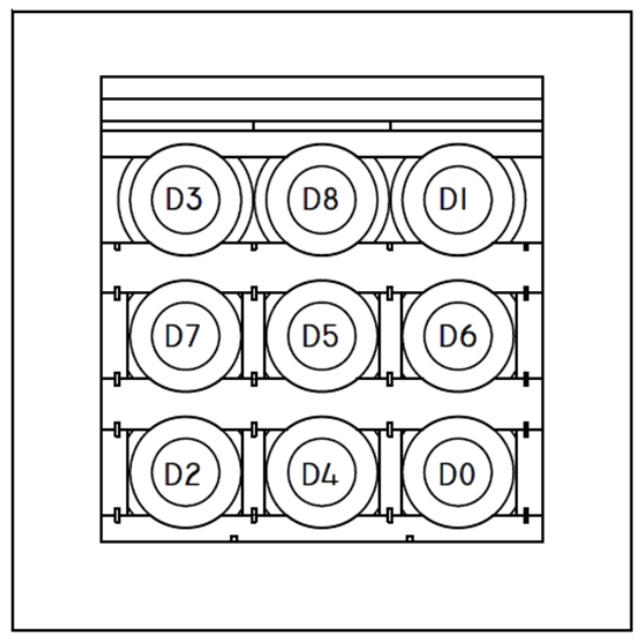}
 \caption{Design of the whole ANAIS-112 set-up at LSC (left) and detector positions in the 3$\times$3 modules matrix (right); detectors are hold in place by a light Teflon structure. The nine detectors are encapsulated in OFE copper and shielded by 10~cm of archaeological lead, 20~cm of low activity lead, 40~cm of neutron moderator combining polyethylene bricks and water tanks, an anti-radon box and an active muon veto system made up of 16 plastic scintillators covering the top and sides of the whole set-up.}
  \label{anais112setup}
\end{figure*}

The shielding for the experiment consists of 10~cm of archaeological lead, 20~cm of low activity lead, 40~cm of neutron moderator, an anti-radon box (continuously flushed with radon-free air) and an active muon veto system made up of plastic scintillators covering the top and sides of the whole set-up. A radon-free calibration system has been implemented to allow for periodic and simultaneous calibration of all the modules at low energy using $^{109}$Cd sources on flexible wires. The hut housing the experiment is placed at the hall B of LSC (under 2450~m.w.e. of overburden). Figure~\ref{anais112setup} shows a design of the set-up and the detector positions inside the shielding.

\begin{table*}
\caption{Powder name and date of arrival at LSC for the nine detectors of ANAIS-112 produced by Alpha Spectra.}
\label{crystals}
\begin{center}
\begin{tabular}{lcc}
\hline\noalign{\smallskip} Detector &  Powder Name &   Date of arrival at Canfranc
 \\ \noalign{\smallskip}\hline\noalign{\smallskip}
D0, D1 & $<$90 ppb K & December 2012 \\
D2 & WIMPScint-II & March 2015 \\
D3 & WIMPScint-III & March 2016 \\
D4, D5 & WIMPScint-III & November 2016 \\
D6, D7, D8 & WIMPScint-III & March 2017
\\ \noalign{\smallskip}\hline
\end{tabular}
\end{center}
\end{table*}

Concerning the data acquisition system (DAQ), each PMT charge output signal is separately processed for obtaining trigger, pulse shape digitization and energy in two ranges: low and high energy. Triggering of every detector is done by the coincidence (logical AND) of the two PMT signals of any detector at photoelectron level in a 200~ns window, while the main acquisition trigger is the logical OR of individual detectors. Electronics is placed at an air-conditioned-room to decouple from temperature fluctuations. The muon detection system using plastic scintillators is integrated in the DAQ in order to monitor on-site the muon flux and to tag muon-related events. The slow control to monitor the relevant  environmental parameters (like radon content, humidity, pressure, different temperatures, N$_{2}$ flux or PMT HV) is also ongoing since the start of ANAIS-112 dark matter run, saving data every few minutes and generating alarm messages when necessary. The building of the spectra is done by software (off-line) by adding the signals from both PMTs, and Pulse Shape Analysis is applied in order to select bulk scintillation events in the NaI crystals and to distinguish alpha interactions from beta/gamma ones.
Filtering protocols for PMT noise similar to those described at \cite{anaisepjc} but updated and optimized for these detectors have been applied; this filtering procedure and the robust acceptance efficiency estimate, described in a companion paper \cite{anaiscompanion}, have allowed to reach 1~keV analysis threshold. In addition, a blank module has been integrated in the set-up at LSC in August 2018 to monitor specifically non-NaI(Tl) scintillation events along the second year of operation. The blank module is similar to the other ones, but without NaI(Tl) crystal; two PMTs, identical to those used in ANAIS-112 modules, have been coupled to quartz optical windows at both sides of the module, consisting of a copper housing with the interior covered by Teflon diffusor film. It is installed in the ANAIS hut, inside a specific lead shielding, besides ANAIS-112; the electronics, DAQ and analysis procedures are the same than for the rest of the modules. It is worth noting that monitoring the muon flux and the PMT noise events is very important to control the systematics in the annual modulation analysis.

Before ANAIS-112 commissioning, some of the modules had been tested and characterized in different smaller but equivalent set-ups: ANAIS--25 \cite{anaisnima} took data with D0 and D1 detectors from December 2012 to March 2015; in ANAIS--37, D2 was added and also D3, D4 and D5 detectors were successively operated in three-module  set-ups with different detector configurations (A37D3, A37D5). From March to July 2017, the commissioning phase of the full ANAIS-112 experiment took place performing calibration and general assessment, and the dark matter run started on 3$^{rd}$ August 2017, and is going on smoothly since then with very high duty cycle and good stability on trigger rate and gain; the performance of ANAIS-112 detectors during the first year of data taking (with a live time of 341.72~days) is described in detail in the companion paper \cite{anaiscompanion}. In the dark matter run the region of interest is blinded except for events in coincidence among detectors, as they  cannot be ascribed to dark matter and are used for tuning the event selection protocols; $\sim$10\% of the total statistics ($\sim$32.9 days of live time selected randomly from the whole first year dark matter data taking) has been unblinded for background assessment and tuning some of the procedures, while the rest of data has remained blinded.

\section{Quantification of background sources} \label{quanti}

The background sources considered for the ANAIS detectors include activity from crystals as well as from external components. Concerning the NaI(Tl) crystals, the activity of the most relevant primordial and cosmogenic radionuclides has been directly measured for our detectors applying different techniques. Background sources from the crystals are covered in subsections \ref{seck} to \ref{sech}. For the other components of the set-up, HPGe spectrometry has been used to determine the relevant radioactive contaminants, in particular PMTs are an important background source in ANAIS-112, as detailed in subsection \ref{ea}. Contribution from muons interacting in the crystals (and other muon related events) can be vetoed by the coincidence with a signal in the plastic scintillators covering the shielding and then, it has not been considered in our background model.  

\subsection{External activity}
\label{ea}

Primordial activity in the components of the set-up other than the crystals has been mainly directly assayed by HPGe spectrometry at LSC. Table~\ref{extcon} summarizes the measured activities (or derived upper limits) taken into consideration in the ANAIS-112 background model. Every PMT unit used in the full experiment has been screened, finding compatible levels of activity among them; values quoted in table~\ref{extcon} correspond to the eighteen units used in the ANAIS-112 set-up, being the weighted mean values also shown. As it will be discussed in section~\ref{comparison}, the radioactivity of PMTs is relevant not only for the detection of their emissions at the NaI(Tl) crystals but also for the induced Cerenkov emission. For copper and quartz windows values are the same as was reported for ANAIS--0 prototype \cite{anaisap}. For the radon content in the air filling the inner volume of the shielding, there is no direct measurement. Radon in the laboratory air is being continuously monitored, and the inner volume of the shielding is flushed with boil-off nitrogen to guarantee its radon-free quality. A value for the radon content in the inner volume air of about one hundredth of the external air radon content has been assumed in our background model (0.6~Bq/m$^{3}$).

\begin{table*}
\caption{Activity of the external components (outside crystal) of
the ANAIS-112 detectors considered as background sources. Except for
the inner volume atmosphere, the values have been measured by HPGe
spectrometry performed at the LSC. Values obtained for each Hamamatsu R12669SEL2 PMT unit are reported together with the weighted mean for all of them. Upper limits are given at 95\%
C.L.}
\label{extcon}
\begin{center}
\begin{tabular}{lcccccc}
\hline\noalign{\smallskip} Component &  Unit  &  $^{40}$K & $^{232}$Th &  $^{238}$U  & $^{226}$Ra & Others
 \\ \noalign{\smallskip}\hline\noalign{\smallskip}

PMTs D0 & mBq/PMT & 	97$\pm$19	&20$\pm$2&	128$\pm$38&	84$\pm$3& \\
	&& 133$\pm$13&	20$\pm$2&	150$\pm$34&	88$\pm$3& \\
PMTs D1	& mBq/PMT & 105$\pm$15&	18$\pm$2&	159$\pm$29&	79$\pm$3& \\
	&&	105$\pm$21&	22$\pm$2&	259$\pm$59&	59$\pm$3&\\
PMTs D2	& mBq/PMT &155$\pm$36&	20$\pm$3&	144$\pm$33&	89$\pm$5&\\
	&&	136$\pm$26&	18$\pm$2&	187$\pm$58&	59$\pm$3& \\
PMTs D3	& mBq/PMT &108$\pm$29&	21$\pm$3&	161$\pm$58&	79$\pm$5&\\
	&&	95$\pm$24&	22$\pm$2&	145$\pm$29&	88$\pm$4&\\
PMTs D4	& mBq/PMT &98$\pm$24&	21$\pm$2&	162$\pm$31&	87$\pm$4&\\
	&&	137$\pm$19&	26$\pm$2&	241$\pm$46&	64$\pm$2&\\
PMTs D5	& mBq/PMT &90$\pm$15&	21$\pm$1&	244$\pm$49	&60$\pm$2&\\
	&&	128$\pm$16&	21$\pm$1&	198$\pm$39&	65$\pm$2&\\
PMTs D6	& mBq/PMT &83$\pm$26&	23$\pm$2&	238$\pm$70&	53$\pm$3&\\
	&&	139$\pm$21&	24$\pm$2&	228$\pm$52&	67$\pm$3&\\
PMTs D7	& mBq/PMT &104$\pm$25&	19$\pm$2&	300$\pm$70&	59$\pm$3&\\
	&&	103$\pm$19&	26$\pm$2&	243$\pm$57	&63$\pm$3&\\
PMTs D8	& mBq/PMT &127$\pm$19&	23$\pm$1&	207$\pm$47&	63$\pm$2&\\
	&&	124$\pm$18&	21$\pm$2&	199$\pm$44&	61$\pm$2&\\
weighted mean & mBq/PMT &	114.9$\pm$4.6&	21.6$\pm$0.4&	180.2$\pm$9.8&	66.7$\pm$0.6&
\\ \hline

Copper encapsulation & mBq/kg & $<$4.9 &  $<$1.8 &  $<$62 & $<$0.9
&$^{60}$Co: $<$0.4 \\ \hline

Quartz windows &  mBq/kg & $<$12 & $<$2.2 & $<$100 & $<$1.9 & \\
\hline

Silicone pads & mBq/kg & $<$181 &  $<$34 & &  51$\pm$7 & \\ \hline

Archaelogical lead & mBq/kg & &  $<$0.3 & $<$0.2 & & $^{210}$Pb:
$<$20 \\ \hline

Inner volume atmosphere & Bq/m$^{3}$ &   & & & &  $^{222}$Rn: 0.6
\\ \noalign{\smallskip}\hline
\end{tabular}
\end{center}
\end{table*}

Contributions from fast neutrons and environmental gamma background have been analyzed and determined to be negligible compared to the present level of sensitivity. The simulated contribution of environmental neutrons at LSC (as measured at \cite{Carmona:2004qk,Jordan:2013} and considering the typical evaporation spectrum) gives a contribution of tenths of counts keV$^{-1}$ kg$^{-1}$ d$^{-1}$ in the region of interest in absence of neutron shielding; the 40 cm of moderator implemented in ANAIS-112 reduce this to a totally negligible level thanks to the suppression of the fast neutron flux. Concerning muon-induced neutrons in the rock surrounding the laboratory, the expected flux is $\sim$3 orders of magnitude lower than that of the environmental neutrons~\cite{Carmona:2004qk}. A simulation of higher energy neutrons considering the production rate of muon-induced neutrons in the lead shielding at LSC from ~\cite{Carmona:2004qk} points to a contribution of $\sim$10$^{-2}$~keV$^{-1}$ kg$^{-1}$ d$^{-1}$ in the region of interest for this type of neutrons without considering the tagging in the veto system. The 30-cm-thick lead shielding allows for a reduction of at least six orders of magnitude of the environmental gamma flux measured for $^{238}$U, $^{232}$Th and $^{40}$K emissions at LSC \cite{bettini2012}.

\subsection{$^{40}$K}
\label{seck}

The measured potassium content of low background NaI(Tl) crystals from different suppliers is of the order of hundreds of ppb \cite{anaisijmpa}, while crystals from DAMA /LIBRA, produced by Saint-Gobain Company, have levels at or below $\sim$20 ppb \cite{bernabei2008}; but the manufacturing procedure used by Saint-Gobain was never disclosed to third parties nor used for other customers.
After contacts and a long-term collaboration with the ANAIS group, detectors produced by the company Alpha Spectra Inc. have shown to have a potassium activity only slightly higher than that of the DAMA/LIBRA crystals. Different powders have been considered and used for the growing of the ANAIS modules (see tables~\ref{crystals} and \ref{intcon}).

\begin{figure*}
\centering
 \includegraphics[width=0.75\textwidth]{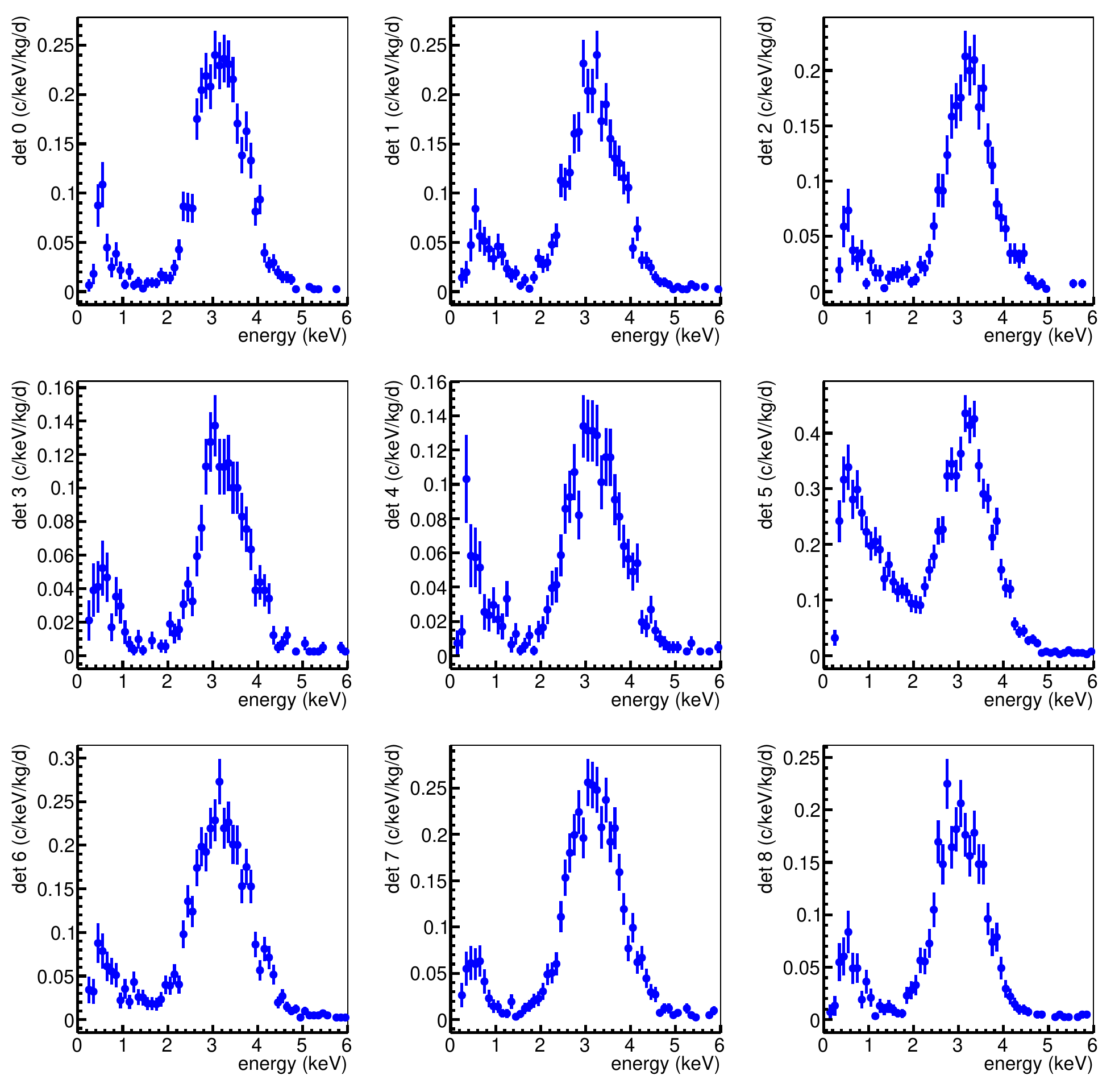} 
 \caption{Low energy spectra of events in coincidence with a 1460.8~keV gamma (in a $\pm$1.4$\sigma$ window) in another module used to assess the $^{40}$K activity in each crystal by quantifying the 3.2~keV emissions. They were obtained with the ANAIS-112 set-up for all the detectors during the first year of data taking, applying the relevant filtering procedures and the corresponding efficiency correction \cite{anaiscompanion}. The peak at 0.3~keV from L-shell electron capture can also be clearly identified.}
  \label{potassium}
\end{figure*}

The bulk $^{40}$K content of ANAIS-112 crystals has been evaluated by searching for the coincidences between 3.2~keV energy deposition in one detector (following the electron capture decay from  K-shell) and the 1460.8~keV gamma line escaping from it and being fully absorbed in other detectors \cite{anaisijmpa}; efficiency of the coincidence is estimated by Monte Carlo simulation. Figure~\ref{potassium} shows the spectra at low energy in coincidence with the high-energy gamma at 1460.8~keV (in a $\pm$1.4$\sigma$ window) in another module, for all the detectors in the ANAIS-112 set-up during the first year of data taking; the relevant filtering procedures and the corresponding efficiency correction \cite{anaiscompanion} have been applied and in these plots the 0.3~keV peak from L-shell electron capture can be clearly identified as well. Table~\ref{intcon} presents the measured values of $^{40}$K activity for the nine crystals from ANAIS-112 data; compatible results were previously found for detectors D0 to D5 using the data from the set-ups with only two or three modules, having lower coincidence probability. On average, the $^{40}$K activity in ANAIS-112 crystals is 0.96~mBq/kg, corresponding to 32~ppb of K. The potassium concentration in the crystals of the COSINE-100 experiment produced by the same company is very similar \cite{cosinebkg}.

\begin{table*}
\caption{Measured activity in NaI(Tl) crystals determined independently for each one of the nine ANAIS-112 detectors combining different analysis techniques. Quantification of $^{232}$Th and $^{238}$U activity by analyzing the alpha events has only been possible for some detectors as in the other cases alpha particles produce saturated signals. Note that the reported $^{210}$Pb activity was determined when detectors arrived at Canfranc; the decay, even if slow, is already noticeable for some detectors in figure~\ref{pb210}.
} \label{intcon}
\begin{center}
\begin{tabular}{lcccc}
\hline\noalign{\smallskip} Detector &   $^{40}$K & $^{232}$Th &  $^{238}$U  & $^{210}$Pb \\
& (mBq/kg) & (mBq/kg) & (mBq/kg)& (mBq/kg)
 \\ \noalign{\smallskip}\hline\noalign{\smallskip}
D0 &  1.33$\pm$0.04  & (4$\pm$1) 10$^{-3}$ &(10$\pm$2) 10$^{-3}$ & 3.15$\pm$0.10 \\
D1 &  1.21$\pm$0.04 & & & 3.15$\pm$0.10 \\
D2 &  1.07$\pm$0.03 & (0.7$\pm$0.1) 10$^{-3}$ & (2.7$\pm$0.2) 10$^{-3}$& 0.7$\pm$0.1 \\
D3 &  0.70$\pm$0.03 & & & 1.8$\pm$0.1 \\
D4 &  0.54$\pm$0.04 & & & 1.8$\pm$0.1 \\
D5 &  1.11$\pm$0.02 & & & 0.78$\pm$0.01 \\
D6 &  0.95$\pm$0.03 & (1.3$\pm$0.1) 10$^{-3}$ & & 0.81$\pm$0.01\\
D7 &  0.96$\pm$0.03 & (1.0$\pm$0.1) 10$^{-3}$ & & 0.80$\pm$0.01\\
D8 &  0.76$\pm$0.02 & (0.4$\pm$0.1) 10$^{-3}$ & & 0.74$\pm$0.01\\
\noalign{\smallskip}\hline
\end{tabular}
\end{center}
\end{table*}


\subsection{$^{210}$Pb}

Pulse Shape Analysis (PSA) in NaI(Tl) scintillators allows to powerfully discriminate the alpha origin energy depositions in the bulk, producing a faster scintillation than that corresponding to beta/gamma/muon events; alpha events can be distinguished by comparing the area and the amplitude of the pulse. Hence, the presence of different isotopes from the $^{232}$Th and $^{238}$U chains has been identified for some detectors (in other cases high energy events from alpha particles produce saturated signals) and their activities have been determined by quantifying also Bi/Po and alpha-alpha sequences, finding in all cases very low values of a few $\mu$Bq/kg (see table~\ref{intcon}).

\begin{figure*}
\centering
 \includegraphics[width=\textwidth]{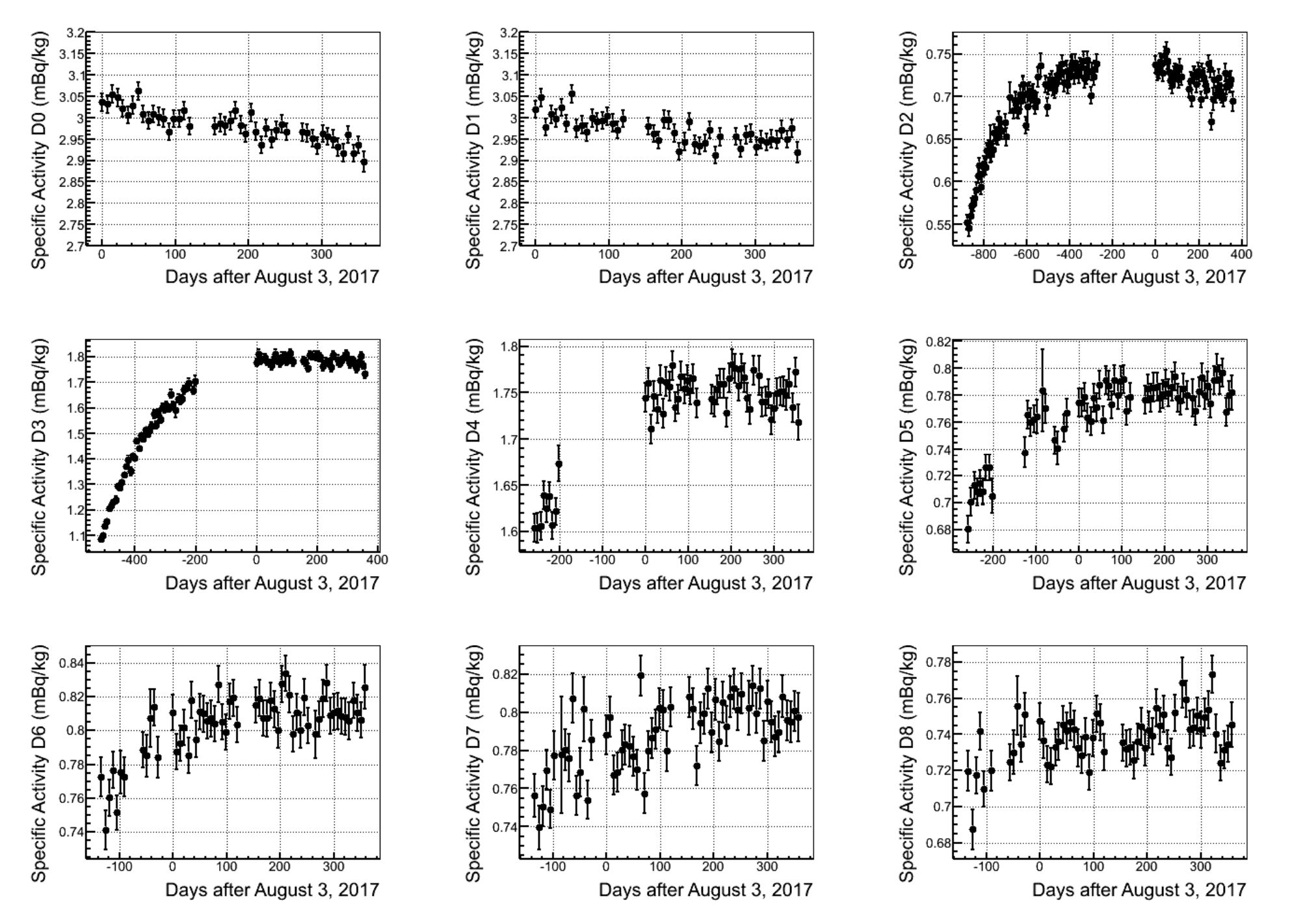}
 \caption{Measured alpha specific activity determined through PSA as a function of time for all the detectors in the ANAIS-112 set-up during the first year of data taking. Activity since the arrival to LSC of each detector is also shown for D2 to D8 detectors.}
  \label{pb210}
\end{figure*}

However, the measured alpha rate following PSA pointed to a relevant activity of $^{210}$Pb out of equilibrium, producing $^{210}$Po.  After the confirmation of the high contamination level of D0 and D1 modules, the possible entrance of radon during growing and/or machining of the detectors was addressed by Alpha Spectra and a significantly lower $^{210}$Pb activity has been measured in the last detectors produced, as shown in table~\ref{intcon}. Nevertheless, these values are still much higher than those achieved in DAMA/LIBRA crystals. Figure~\ref{pb210} shows the time evolution of the alpha activity for all ANAIS-112 detectors along the first year of data taking together with previous results for some detectors; the clear increase observed in some cases is compatible with a $^{210}$Pb contamination at the end of the purification process and growing of the crystal and the subsequent building of $^{210}$Po activity until the equilibrium in the chain is established. Equilibrium has been reached in all detectors and hence, activity from $^{210}$Po should be equal to that from $^{210}$Pb. The $^{210}$Pb activity values presented in table~\ref{intcon} are those determined in the characterization performed when detectors arrived to Canfranc; the decay, even if slow, is already noticeable for the first detectors D0 and D1 (see the corresponding plots of figure~\ref{pb210}).

As for $^{40}$K, the out-of-equilibrium $^{210}$Pb activities in ANAIS-112 crystals are similar to those determined for COSINE-100 crystals produced from the same starting material. It is worth noting that, as it will be shown in section~\ref{comparison}, the assumed $^{210}$Pb activities from the total alpha rate determined through PSA are fully compatible with the measured low energy depositions attributable to $^{210}$Pb.

\subsection{I and Te isotopes}

Several cosmogenically induced isotopes have been identified in ANAIS crystals (see table~\ref{iso}) thanks to different distinctive signatures. The initial specific activities measured when the detectors were moved underground have been determined and, whenever possible, an estimate of the production rates has been made from those specific activities according to the history of detectors. Thanks to the very good detector response and the prompt data taking starting after storing the detectors underground in all ANAIS set-ups, a careful quantification of cosmogenic radionuclide production in NaI(Tl) has been possible. When required, the detection efficiencies for the signals have been obtained by Geant4 simulation taking into account the geometry of each set-up and considering isotopes homogeneously distributed in the crystals.

\begin{table*}
\begin{center}
\caption{Main cosmogenically induced isotopes identified in the data of ANAIS modules. Half-lives and decay mechanisms are indicated. Fourth column shows the energy of analyzed signals: gamma emissions, excited levels of metastable states and binding energies of K-shell electrons for EC decays. Decay information for $^{125}$I, $^{127m}$Te, $^{123m}$Te, $^{113}$Sn, $^{109}$Cd, $^{22}$Na and $^{3}$H has been taken from Ref.~\cite{dec1} and for the other isotopes from Ref.~\cite{dec2}. The last column presents the production rates $R_{p}$ at sea level derived for each isotope from ANAIS data (see text) \cite{anaisjcap}.}
{\begin{tabular}{@{}lcccc@{}} \hline 
 Isotope & Half-life & Decay mechanism & Main $\gamma$ emissions/ &  $R_{p}$ \\
 &&& Metastable levels/  & (kg$^{-1}$d$^{-1}$) \\
 & & & Electron binding energy (keV) &\\
\noalign{\smallskip}\hline\noalign{\smallskip}
$^{126}$I & (12.93 $\pm$ 0.05) d & EC, $\beta^{+}$, $\beta^{-}$ & 666.3 &  283$\pm$36\\
$^{125}$I & (59.407 $\pm$ 0.009) d & EC & 35.5 (+ 31.8) &  220$\pm$10 \\
$^{127m}$Te & (107 $\pm$ 4) d & IT, $\beta^{-}$ & 88.3  & 10.2$\pm$0.4 \\
$^{125m}$Te & (57.4 $\pm$ 0.2) d & IT & 144.8  & 28.2$\pm$1.3 \\
$^{123m}$Te & (119.3 $\pm$ 0.2) d & IT & 247.6  & 31.6$\pm$1.1 \\
$^{121m}$Te & (154 $\pm$ 7) d & IT, EC & 294.0  & 23.5$\pm$0.8 \\
$^{121}$Te & (19.16 $\pm$ 0.05) d & EC & 507.6, 573.1 &  9.9$\pm$3.7\\
$^{113}$Sn & (115.09 $\pm$ 0.03) d & EC & 27.9  & 4.53$\pm$0.40 \\
$^{109}$Cd & (461.9 $\pm$ 0.4) d & EC & 25.5 &  2.38$\pm$0.20\\
$^{22}$Na & (2.6029 $\pm$ 0.0008) y & EC, $\beta^{+}$ & 511, 1274.5 &  45.1$\pm$1.9\\
$^{3}$H & (12.312 $\pm$ 0.025) y & $\beta^{-}$ & &\\
\noalign{\smallskip}\hline
\end{tabular} \label{iso}}
\end{center}
\end{table*}

For I and Te isotopes, initial activities underground were measured following the evolution of the identifying signatures for each isotope along several months. This detailed analysis was made for D0 and D1 detectors in ANAIS-25 set-up; activation has been assessed independently for each crystal and results have been properly combined (full description of the procedure can be found at \cite{anaisjcap}). The evolution in time of the counting rates for the different identifying signatures matched single exponential decays following the radioactive decay law and the half-lives deduced from the fits agreed with the known values within uncertainties, confirming the identification for each isotope. A special analysis was required for $^{121}$Te \cite{anaisjcap}, as it is additionally produced by the decay of $^{121m}$Te, also cosmogenically induced and having a longer half-life.

As indicated before, ANAIS crystals were built in Alpha Spectra facilities in Grand Junction, Colorado. The exposure history of the starting material used to produce the NaI powder from which crystals are grown was not fully known, but for the considered I and Te isotopes saturation must have been reached nonetheless while purifying, growing the crystals and building the detectors, independently from previous exposure history. Colorado is placed at a quite high altitude; consequently, an important correction factor $f$ to the cosmic neutron flux at New York City coordinates has to be taken into consideration. Afterwards, detectors were transported to LSC by boat and by road, being the travel duration about one month, most of the time at sea level. Therefore, the initial activities underground derived here cannot be directly considered the saturation activities (i.e. production rates, $R_{p}$) at sea level. A method to estimate the production rates in our crystals, based on a few reasonable assumptions, was developed \cite{anaisjcap} considering that if saturation activity is reached at a given place characterized by a correction factor $f$ and if the material is then exposed to cosmic rays at sea level for a time $t$, the evolution of the corresponding activity $A(t)$ is:

\begin{equation}
A(t) = R_{p}[1 + (f - 1) exp(-\lambda t)]
\label{nsat}
\end{equation}

The production rate $R_{p}$ of an isotope was obtained from Eq.(\ref{nsat}) using its initial activity underground, and fixing the correction factor $f$ and exposure time at sea level $t$. The correction factor was estimated specifically for Grand Junction altitude~\cite{anaisjcap} as $f =3.6\pm0.1$ and an exposure time of $t=30\pm5$~days was considered. Table~\ref{iso} summarizes the productions rates obtained for I and Te isotopes, used for all ANAIS detectors when including the cosmogenic activation in the corresponding background models; they allow to reproduce successfully the measured spectra for the first data taken underground when activation is very relevant.


Together with these relatively short-lived I isotopes, $^{129}$I can be present in the NaI crystals, produced by uranium spontaneous fission and by cosmic rays. Its concentration is strongly affected by the ore material exposure either to cosmic rays or to high uranium content environment. It presents 100$\%$ $\beta^{-}$ decay to the excited level of 39.6~keV of the daughter nucleus with a half-life T$_{1/2}$~=~(16.1 $\pm$ 0.7)$\cdot$10$^{6}$~y~\cite{dec1}, being hence the expected signature in large NaI(Tl) crystals a continuous beta spectrum starting in 39.6~keV. This signal is above the RoI for dark matter searches, but it is important for a complete understanding of the background. The long lifetime and the difficulty to disentangle the signature from other emissions are the reasons why the quantification of the amount of $^{129}$I in ANAIS crystals was not possible. In Ref.~\cite{bernabei2008} the estimated fraction of this isotope was determined to be $^{129}$I/$^{nat}$I = (1.7 $\pm$ 0.1)$\cdot$10$^{-13}$. A first measurement at Centro Nacional de Aceleradores (Sevilla, Spain) in order to measure the real content in this isotope for ANAIS crystals using Alpha Spectra NaI powder has given an upper limit set by Accelerator Mass Spectrometry that is a factor 2.5 higher than DAMA/LIBRA value. To take $^{129}$I into account in the ANAIS background models, its concentration was assumed to be the same as estimated by DAMA/LIBRA, corresponding to an activity of 0.94~mBq/kg.


\subsection{$^{109}$Cd and $^{113}$Sn}
\label{seccdsn}

As summarized in table~\ref{iso}, $^{113}$Sn decays by electron capture mainly to a 391.7~keV isomeric state of the daughter, having a half-life of 115.1~days; therefore, a peak at the binding energy of K-shell electrons of In at 27.9~keV is produced, as well as another one for the L-shell at around 4~keV. The ratio between the probabilities of electron capture for K and L shells is 7.4 for decay to the isomeric state \cite{dec1}. Similarly, $^{109}$Cd decays by electron capture to the 88-keV isomeric state of the daughter, having a half-life of 461.9~days, and therefore it gives a peak at the binding energy of the K-shell of Ag at 25.5~keV together with an additional peak around 3.5~keV, corresponding to the Ag L-shell binding energy (being 5.4 the ratio between K and L-shell EC probabilities \cite{dec1}).

\begin{figure*}
\centering
\includegraphics[width=.45\textwidth]{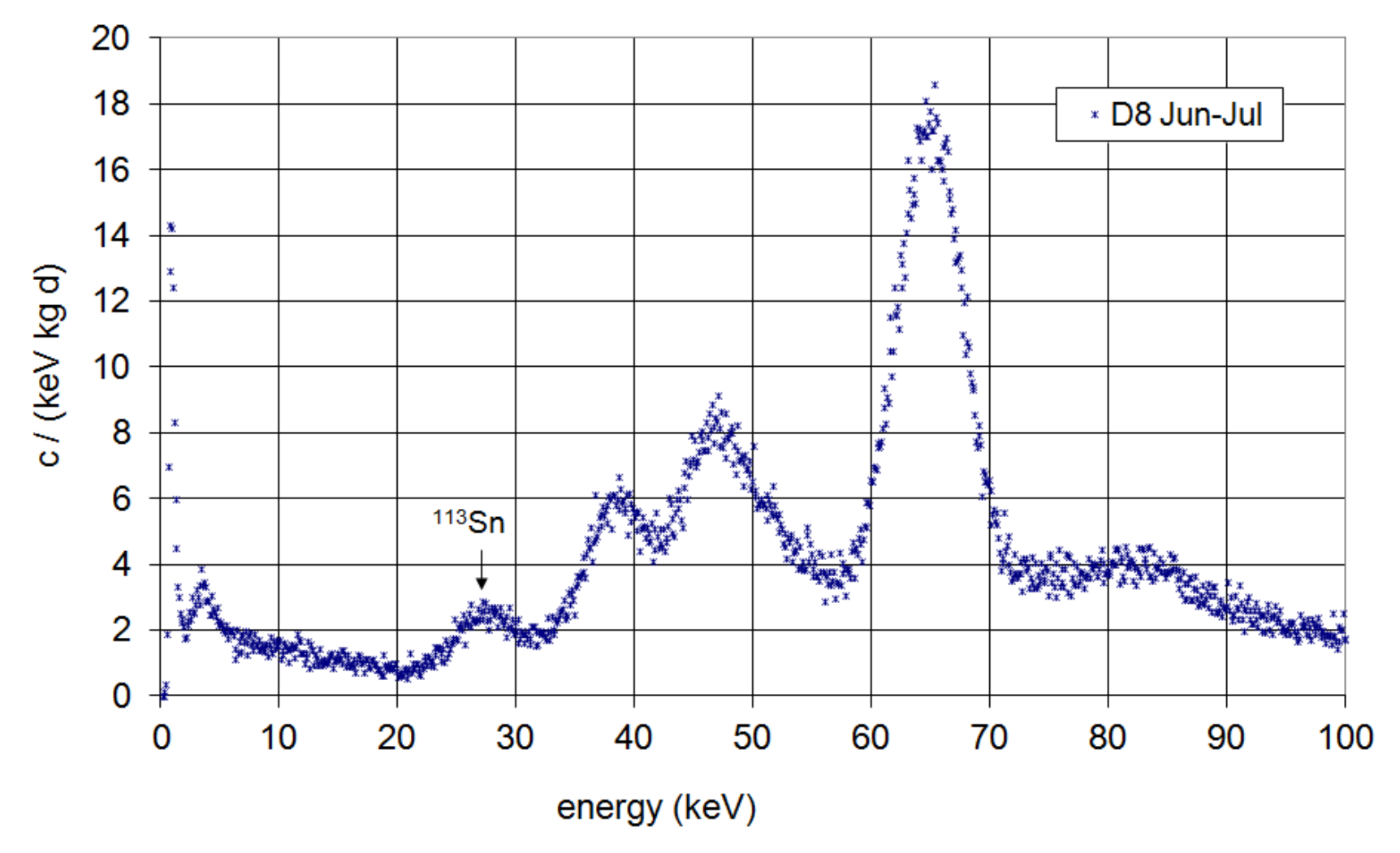} 
\includegraphics[width=.43\textwidth]{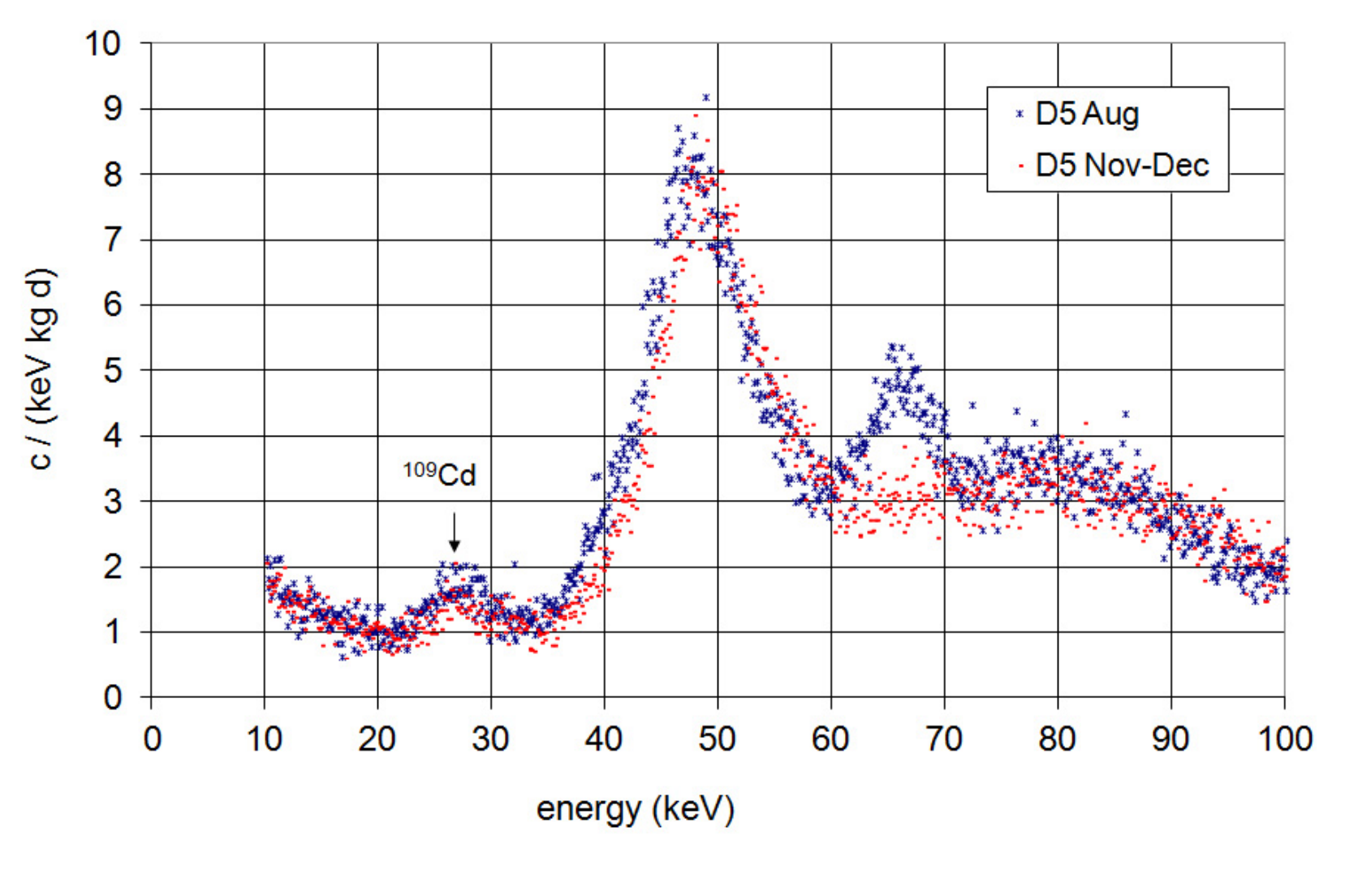}
 \caption{Examples of the energy spectra considered in the quantification of $^{113}$Sn and $^{109}$Cd signals from the ANAIS-112 data: spectrum of D8 detector taken during June-July 2017 after being $\sim$4 months underground used for $^{113}$Sn (left) and those of D5 detector during August and November-December in 2017 after $\sim$9 and 12 months underground used for $^{109}$Cd (right). Very similar spectra are registered for equivalent detectors and runs. Note that the dark matter run of ANAIS-112 started on 3$^{rd}$ August 2017 and data of the very low energy region are blinded since then.}
\label{SpcSnCd}
\end{figure*}

In the first analysis of cosmogenic activation carried out using ANAIS-25 data \cite{anaisjcap}, the presence of $^{113}$Sn and $^{109}$Cd in the NaI(Tl) crystals was not observed; however, the study of the background measured in D0, D1 and D2 detectors performed in a longer term allowed to identify signals from these two isotopes~\cite{anaisbkg}. For the ANAIS detectors operated in the different set-ups, the following behavior has been observed: once the activities of tellurium and iodine cosmogenic isotopes decaying by EC have significantly decreased, the large peak produced at the K-shell binding energies of Sb and Te (see plots in figure~\ref{SpcSnCd} as an example) disappears and a line at $\sim$28~keV can be identified, which also fades away within a few months; then, the contribution at 25~keV is visible, decreasing in time more slowly. Although a precise time analysis cannot be attempted as the peaks are very small, even if accumulated over long times, and cannot be resolved separately, the observed behavior seems compatible with the hypothesis that those peaks can be attributed to the decays of $^{113}$Sn and $^{109}$Cd cosmogenically produced in the NaI(Tl) crystals. Peaks around 3.5 and 4~keV from $^{109}$Cd and $^{113}$Sn decays, corresponding to the binding energies of the L-shells of Ag and In, respectively, are expected in the RoI for dark matter searches. Therefore, these isotopes must be taken into account in the background models of NaI(Tl) detectors; as the half-lives, especially for $^{113}$Sn, are not too large, they should not be a problem in the long term.

A quantification of the induced activities and the corresponding production rates of these two isotopes has been attempted profiting from data taken in the ANAIS crystals along different times, analyzing the signals registered at the K-shell binding energies (with probabilities 0.8374 for $^{113}$Sn and 0.8120 for $^{109}$Cd, according to the decay data from \cite{dec1}) accumulated over a quite long time to make them measurable (see figure~\ref{SpcSnCd}); the whole area of the peaks has been assumed to be due just to one of the isotopes in the analyzed runs. This overestimates somehow the derived activities. Then, the induced activity is deduced and, from the assumed exposure history of the detectors, the production rates can be estimated, as made for I and Te isotopes \cite{anaisjcap}. It is worth noting that the signatures to identify $^{113}$Sn and $^{109}$Cd were already imperceptible in all the data analyzed here for detectors D0 and D1, which have been underground for a longer time. Table~\ref{SnCd} presents the initial activities obtained for several detectors and the deduced production rates, which being mostly compatible within uncertainties, have been averaged because the exposure history of the detectors is supposed to be quite similar for most of them, as it will be further discussed in section~\ref{secna}. The weighted mean rates, also reproduced in table~\ref{iso}, have been properly taken into account in the background models of the nine detectors; these emissions are essential to reproduce some features of the measured low energy spectra. The productions rates obtained in our study are around a factor of two lower than those estimated by DM-Ice \cite{dmicethesis}.

\begin{table*}
\begin{center}
\caption{Initial activities A$_{0}$ and production rates at sea level R$_{p}$ for
$^{113}$Sn and $^{109}$Cd deduced from data of several detectors in different ANAIS set-ups. All results are expressed in kg$^{-1}$d$^{-1}$.}
\begin{tabular}{@{}lcccc@{}}
\hline 
Detector &  A$_{0}$ & A$_{0}$ &  R$_{p}$  &  R$_{p}$ \\
& $^{109}$Cd &  $^{113}$Sn & $^{109}$Cd & $^{113}$Sn \\
\hline 
D2 & 11.7$\pm$2.4 & 18.7$\pm$4.4 & 3.37$\pm$0.71 & 5.89$\pm$1.41 \\
D3 & 6.5$\pm$2.1 & 11.5$\pm$4.2 & 1.86$\pm$0.60 & 3.64$\pm$1.33 \\
D4 & 10.0$\pm$2.7 & 20.5$\pm$6.2 & 2.88$\pm$0.77 & 6.47$\pm$1.96 \\
D5 & 9.2$\pm$2.3 & 19.2$\pm$4.6 & 2.63$\pm$0.65 & 6.06$\pm$1.45 \\
D6 & 8.9$\pm$1.6 & 14.8$\pm$2.9 & 2.54$\pm$0.46 & 4.66$\pm$0.91 \\
D7 & 7.9$\pm$1.4 & 13.4$\pm$2.5 & 2.26$\pm$0.40 & 4.22$\pm$0.80 \\
D8 & 7.2$\pm$1.4 & 12.2$\pm$2.5 & 2.08$\pm$0.39 & 3.85$\pm$0.79 \\ \hline
Weighted mean & & & 2.38$\pm$0.20 & 4.53$\pm$0.40 \\
\hline 
\end{tabular} \label{SnCd}
\end{center}
\end{table*}

\subsection{$^{22}$Na}
\label{secna}

$^{22}$Na is specially worrisome for dark matter searches because the binding energy of the K-shell of its daughter Ne is 0.87~keV, falling the corresponding energy deposition in the RoI, and having a long enough half-life to compromise the first years of data taking. The initial activity in crystals when moving underground has been estimated for the nine modules used in ANAIS-112, being different as the exposure history was not the same for all the detectors. D0 and D1 seem to have had a longer exposure than the other detectors. In addition, a shelter for cosmic rays was used during production for some of the detectors.

In detectors D0 and D1, cosmogenically activated $^{22}$Na was firstly identified and quantified using coincidence spectra in data from March 2014 to June 2014 in ANAIS-25. The analyzed signature was the integral number of events from 2300 to 2900~keV (corresponding to full absorption of its positron and gamma emissions) in the spectrum obtained summing D0 and D1 energies, for coincidence events leaving 511 (or 1274.5)~keV at any detector. These were the best signatures found for this isotope to avoid the interference of backgrounds.
The effect of coincidences not due to $^{22}$Na was taken into account subtracting their contribution in the 2300-2900~keV region, evaluated from events in a window on the left of the 511~keV peak and on the right of the 1274.5~keV peak. From the net number of events registered in the 2300-2900 keV region and the estimated efficiency the activity $A$ of $^{22}$Na at measuring time was evaluated and then the initial activity underground A$_{0}$ deduced, for both 511 and 1274.5 keV~windows; compatible results were found and the average value is presented in table~\ref{tana22}. From the $A_{0}$ results and taking into account the exposure history of the detectors, the production rate of $^{22}$Na at sea level was also evaluated \cite{anaisjcap} and it is presented in table~\ref{iso}.

A direct estimate of $^{22}$Na activity in D2 crystal was carried out for the first time by analyzing coincidences from data corresponding to 111.4 days (live time) from a special set-up of ANAIS-37 taking data from October 2015 to February 2016 where only D0 and D2 detectors were used; in particular, profiting from the reduced cosmogenic contribution to the background in this period after enough cooling time underground, D2 spectrum in coincidence with 1274.5~keV depositions in D0 was analyzed. The obtained value for the initial activity (corresponding to the moment of storing crystals deep underground at LSC), A$_{0}$ = (70.2 $\pm$ 3.9) kg$^{-1}$ d$^{-1}$, was more than a factor of two lower than the one deduced for D0 and D1 detectors~\cite{anaisepjc}. This result was compatible with a lower time of exposure of D2 to cosmic rays, taking into account the $^{22}$Na half-life, longer than that corresponding to I and Te products. It is worth noting that the $^{22}$Na initial activity in D0 deduced from the analogue analysis is in perfect agreement with the first estimate in ANAIS-25 set-up (see table~\ref{tana22}).

A procedure to quantify the initial activities  of $^{22}$Na in the rest of detectors analogue to that followed for D2 has been applied in different set-ups with three crystals and in ANAIS-112 with nine crystals, looking for coincidences with 1274.5~keV depositions in other detectors. In particular, data from August to November 2017 collected in ANAIS-112 have been considered \cite{ijmpaNaI}. Preliminary results from ANAIS-112 together with all the available previous estimates are summarized in table~\ref{tana22}. The initial activity for D0 and D1 was significantly larger than for the other detectors produced afterwards, pointing at a reduction on the exposure while crystal growing and detector building at Colorado. Whenever estimates from different set-ups are available, a reasonable agreement between them is obtained except for D2.

It is worth noting that the $^{22}$Na activity measured in ANAIS-112 crystals is very similar to that found in the crystals of the COSINE-100 experiment produced by the same company \cite{cosinebkg}. It is also at the same order of the value measured by the SABRE experiment using a HPGe at the Gran Sasso Laboratory for AstroGrade quality NaI powder from Sigma-Aldrich, being 0.48~mBq/kg $=$ 41.5~kg$^{-1}$ d$^{-1}$ \cite{sabrebkg}. The similarity of these $^{22}$Na activities points to a comparable exposure history, and therefore, similar activity for other cosmogenic isotopes would be also expected.

\begin{table*}
\begin{center}
\caption{Comparison of cosmogenically produced $^{22}$Na initial (when moving detectors underground) activity, $A_{0}$, estimated (in kg$^{-1}$ d$^{-1}$) for ANAIS detectors in different set-ups (see text).}
\begin{tabular}{@{}ccccc@{}} \hline 
Detector & ANAIS-25~\cite{anaisjcap} & ANAIS-37~\cite{anaisbkg} & A37D3 & ANAIS-112  \\
 \noalign{\smallskip}\hline\noalign{\smallskip}
D0 & 159.7 $\pm$ 4.9 & 158.4 $\pm$ 7.9 & 164 $\pm$ 17 & 155 $\pm$ 11\\
D1 & 159.7 $\pm$ 4.9 &  &  & 168 $\pm$ 11\\
D2 &  & 70.2 $\pm$ 3.9 & 57.6 $\pm$ 8.1 & 43.9 $\pm$ 6.0  \\
D3 &  &  & 69.9 $\pm$ 3.6 & 68.6 $\pm$ 4.6  \\
D4 & & &  & 61.8 $\pm$ 3.1\\
D5 &  &  & & 43.7 $\pm$ 2.3 \\
D6 &  &  & & 53.8 $\pm$ 2.7 \\
D7 &  &  & & 55.6 $\pm$ 2.7 \\
D8 &  &  &  & 56.4 $\pm$ 2.8\\
\noalign{\smallskip}\hline\noalign{\smallskip}
\end{tabular} \label{tana22}
\end{center}
\end{table*}

\begin{figure*}
\centering
 \includegraphics[width=1\textwidth]{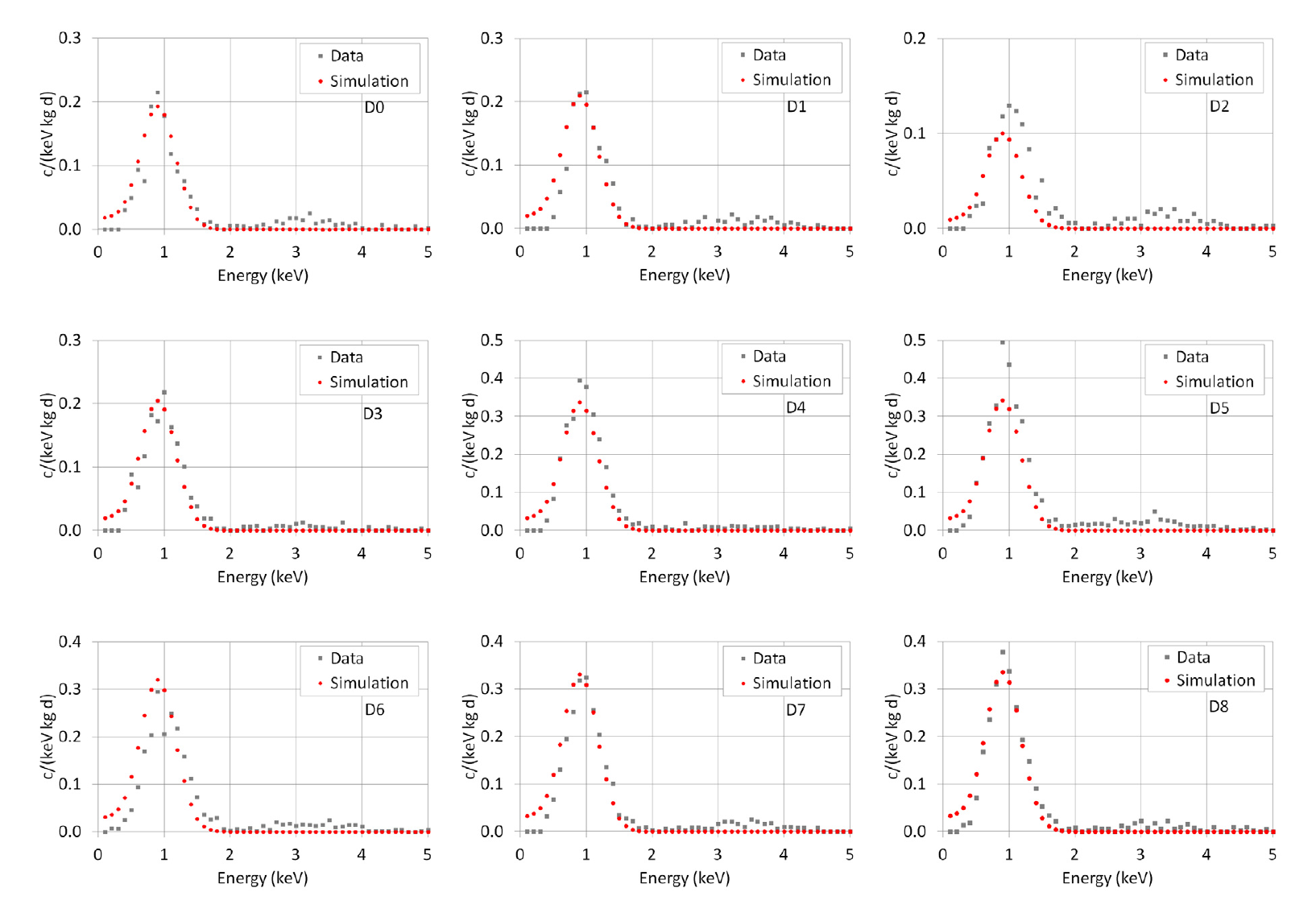}
 \caption{Comparison for the nine ANAIS-112 detectors in the first year of data taking of the low energy spectra (after filtering and efficiency correction) in coincidence with the high energy gamma at 1274.5~keV with the corresponding simulation using the $^{22}$Na activity values deduced from different high energy signatures.}
  \label{Nacoin}
\end{figure*}

The population of coincidence events having an energy deposition of 0.87~keV in one detector (following the electron capture decay of $^{22}$Na) and the 1274.5~keV gamma line, escaping from it and being fully absorbed in other detector, offers the possibility to cross-check the $^{22}$Na activity determination. The measured rate at 0.87~keV (after filtering and efficiency correction) from the first year of data taking for each ANAIS-112 detector, shown in figure~\ref{Nacoin}, is well reproduced by the corresponding simulation using the deduced $^{22}$Na activity values, calculated using a different signature of the decay, as explained above, and reported in table~\ref{tana22}. Table~\ref{ratena} compares for each detector the measured rate up to 2~keV with the simulated one; for the average rate of all detectors, the deviation is of 0.5\%. In addition, the rate of events at 0.87~keV selected through the coincidence with the high energy gamma has been checked to decay with T$_{1/2} = $(2.7$\pm$0.9)~y \cite{anaiscompanion}, in very good agreement with the half-life of $^{22}$Na. The good description achieved for the coincident 0.87~keV peak allows also to validate the PMTs noise rejection methods and calculated efficiencies even below 1~keV \cite{anaiscompanion}.

\begin{table*}
\begin{center}
\caption{Measured and simulated rates up to 2~keV from spectra shown in figure~\ref{Nacoin} and the corresponding deviation for each ANAIS-112 detector and on average. The statistical uncertainty of simulated rates (not shown in the table) is $\sim$0.1\%.}
\begin{tabular}{@{}cccc@{}} \hline 
Detector & Measurement & Simulation & Deviation  \\
& (kg$^{-1}$ d$^{-1}$) & (kg$^{-1}$ d$^{-1}$) & (\%)  \\
 \noalign{\smallskip}\hline\noalign{\smallskip}
D0 & 0.124$\pm$0.005 & 0.136 & 10.2 \\
D1 & 0.132$\pm$0.006 & 0.148 & 12.4 \\
D2 & 0.095$\pm$0.005 &  0.071 &  -25.1 \\
D3 & 0.141$\pm$0.006 & 0.145 & 2.5 \\
D4 & 0.256$\pm$0.008 & 0.238 & -7.0\\
D5 &  0.296$\pm$0.008 & 0.242 & -18.2  \\
D6 &  0.197$\pm$0.007 & 0.226 & 14.9 \\
D7 &  0.211$\pm$0.007 & 0.234 & 11.0 \\
D8 &  0.235$\pm$0.007 & 0.237 &  1.1\\ \hline
ANAIS-112 & 0.187$\pm$0.007 & 0.186 & -0.5 \\
\noalign{\smallskip}\hline\noalign{\smallskip}
\end{tabular} \label{ratena}
\end{center}
\end{table*}

\subsection{$^{3}$H}
\label{sech}

Tritium is a relevant background in the RoI for the dark matter signal. It is a pure beta emitter with transition energy of 18.591~keV and a long half-life of 12.312~y. Following the shape of the beta spectrum for the super-allowed transition of $^{3}$H, 57$\%$ of the emitted electrons are in the range from 1 to 7~keV; these electrons are typically fully absorbed since most of the dark matter detectors are large enough. Due to the long half-life of tritium, saturation activity is difficult to reach; however, even below saturation, as tritium emissions are concentrated in the energy region where the dark matter signal is expected, tritium can be important. Quantification of tritium cosmogenic production is not easy, neither experimentally since its beta emissions are hard to disentangle from other background contributions, nor by calculations, as tritium can be produced by different reaction channels. Tritium production in materials of interest for dark matter experiments has been studied in Ref.~\cite{tritio}.

\begin{figure*}
\centering
\includegraphics[width=8.3cm]{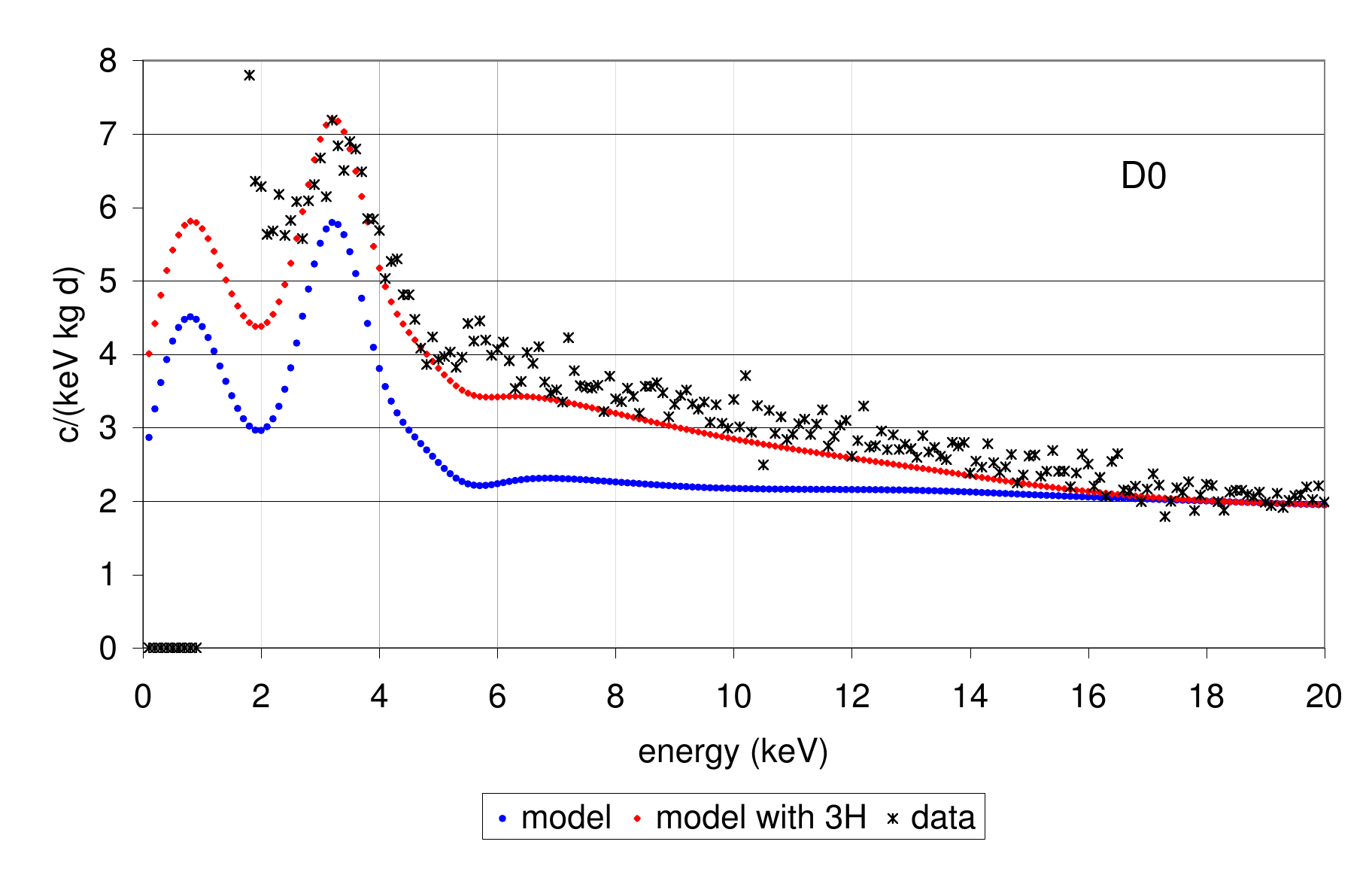}
\includegraphics[width=8.3cm]{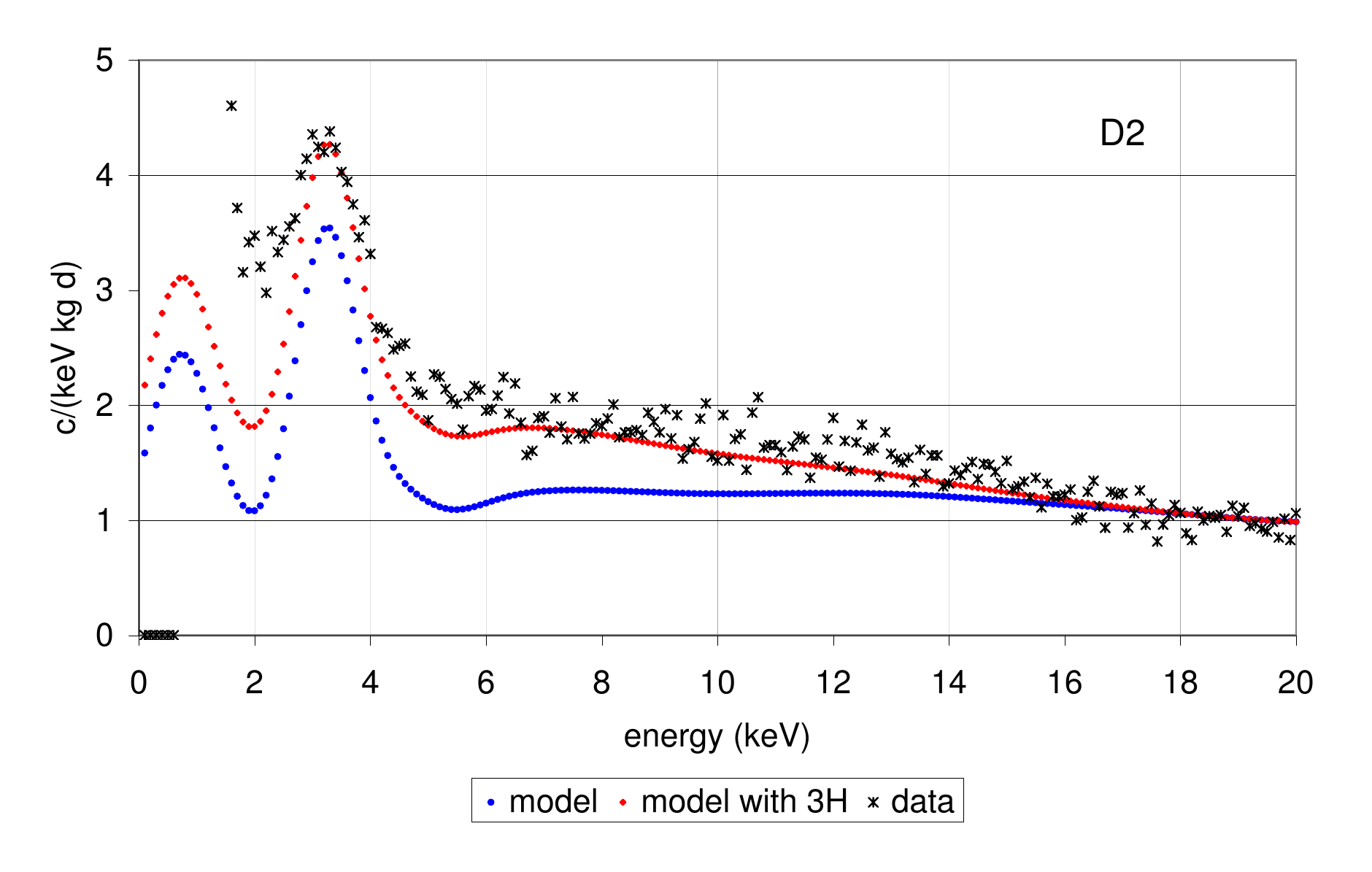}
\includegraphics[width=8.3cm]{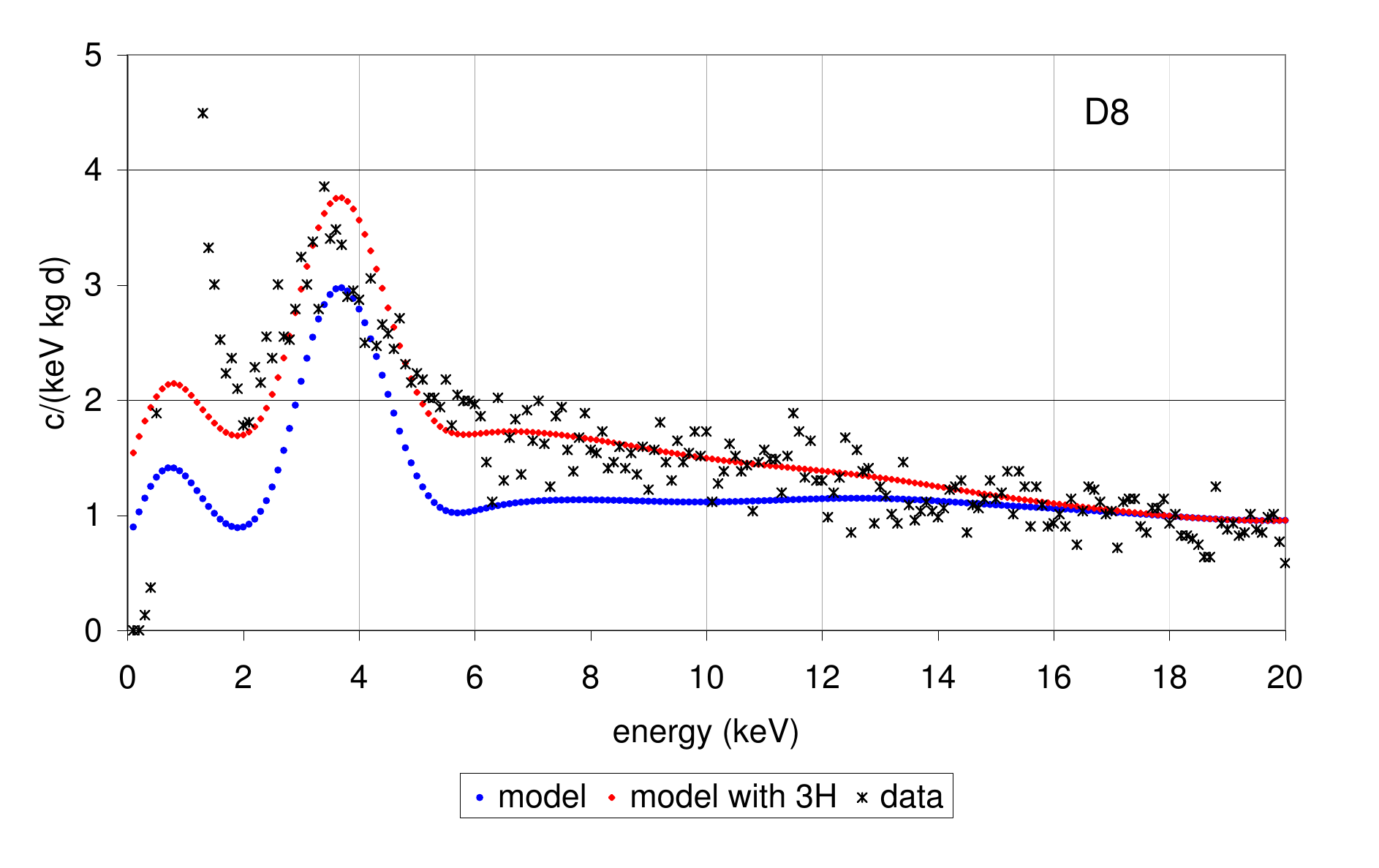}
\caption{The very low energy region of the energy spectra measured for D0 (top, left), D2 (top, right) and D8 (bottom) ANAIS detectors compared with the corresponding simulated models~\cite{anaisbkg} including all the quantified intrinsic and cosmogenic activities in the detectors and main components of the set-up (blue) and adding also tritium in the NaI crystal (red). D0 and D2 data correspond to September - October 2016 in A37D3 set-up and D8 data corresponds to commissioning run of ANAIS-112 from June - July 2017. A $^{3}$H activity of 0.20~mBq/kg is considered for D0 and 0.09~mBq/kg for D2 and D8.}
\label{anais3h}
\end{figure*}

In the ANAIS detectors, the presence of tritium is inferred in order to explain the differences between the measured background and the background models. Although a direct identification of a tritium content in the crystals has not been possible, the construction of detailed background models of the firstly produced detectors pointed to the presence of an additional background source contributing only in the very low energy region, which could be tritium~\cite{anaisbkg,pvillar}. The simulated spectra including all well-known contributions agree reasonably with the ones measured, except for the very low energy region; the inclusion of a certain activity of $^{3}$H homogeneously distributed in the NaI crystal provides a very good agreement also below~20 keV. Figure~\ref{anais3h} compares data and background models for three detectors (D0, D2 and D8; similar results are obtained for the other ones). The required $^{3}$H initial activities to reproduce the data are around 0.20~mBq/kg for D0 and D1 and 0.09~mBq/kg for D2 to D8.

Since the exposure history of the NaI material used to produce the crystals (following different procedures for purification and crystal growth) is not precisely known, no attempt of deriving tritium production rates from these estimated activities in ANAIS crystals has been made. However, the plausibility of the tritium hypothesis has been analyzed: assuming that the whole crystal activation took place in Alpha Spectra facilities at Grand Junction, Colorado (where the cosmic neutron flux is estimated to be a factor $f$ = 3.6 times higher than at sea level~\cite{anaisjcap}) the required exposure time $t_{exp}$ to produce an activity $A$ of an isotope with decay constant $\lambda$ for a production rate $R$ at sea level can be deduced using
\begin{equation}
A = f \cdot R[1-exp(-\lambda \cdot t_{exp})]
\label{activity}
\end{equation}
In Ref.~\cite{tritio}, an estimate of the tritium production rate in NaI of (83$\pm$27)~kg$^{-1}$d$^{-1}$ was obtained, by convoluting a selected description of the excitation functions (based on the TENDL~\cite{tendl} and HEAD~\cite{head} libraries) with the cosmic neutron spectrum at sea level (as parameterized in \cite{gordon})\footnote{The methodology applied is validated by reproducing satisfactorily the measured production rated of tritium in natural germanium by EDELWEISS and CDMSlite experiments (see Ref.~\cite{tritio}).}. For the range of this estimated production rate and the deduced tritium activities in D0 and D2, the exposure times are between 0.8 and 1.6~years and 4.2 and 8.4~months, respectively. These values roughly agree with the time lapse between sodium iodide raw material purification starting and detector shipment, according to the company. As an additional check, for these exposure times, the ratio of the induced initial activities of the also long-living cosmogenic isotope $^{22}$Na in D0 and D2 following Eq.(\ref{activity}) is $\sim$2, in good agreement with the measured activities (see table~\ref{tana22}). Since purification methods cannot remove this isotope, this means that the raw material was not significantly exposed to cosmic rays before.

The $^{3}$H activity estimated for ANAIS-112 crystals is of the same order of that found in the crystals of the COSINE-100 experiment produced by the same company \cite{cosinebkg}. The value for D2-D8 detectors is just the upper limit set for DAMA/LIBRA crystals~\cite{bernabei2008}.

\section{Background simulations} \label{modeling}
\begin{figure*}
\centering
 \includegraphics[height=0.4\textheight]{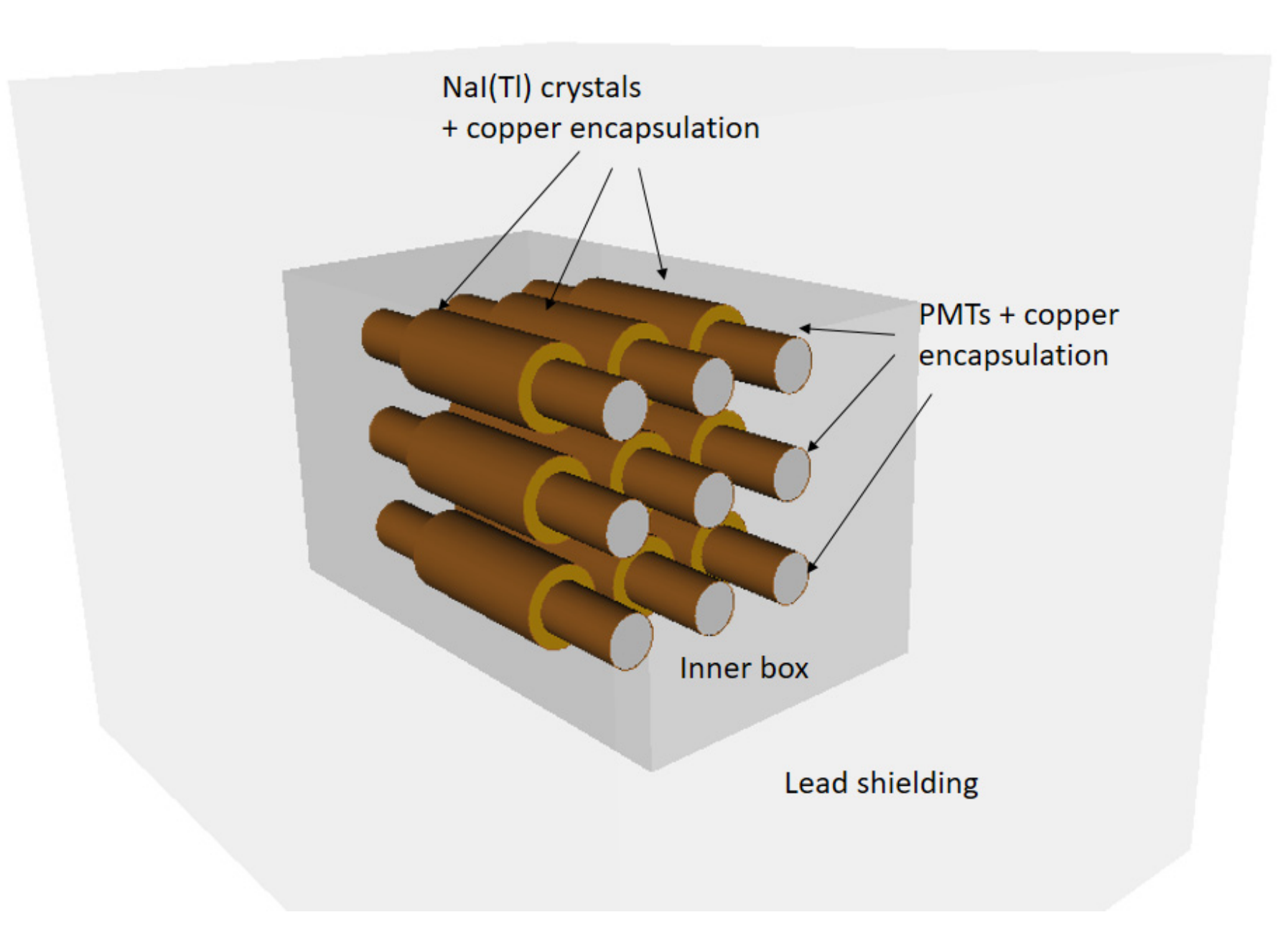}
 \caption{Geometry of the ANAIS--112 set-up implemented in the Geant4 simulations.}
  \label{geometrys}
\end{figure*}

The contribution of all the background sources described in section \ref{quanti} to the background levels of the ANAIS-112 detectors has been simulated by Monte Carlo using the Geant4 package \cite{geant4}, as done in \cite{anaisbkg}. A detailed description of the set-up was implemented including the lead shielding and detectors, considering NaI crystal, teflon wrapping, copper encapsulation with the Mylar window, silicone pads, quartz windows, PMTs, bases and copper enclosure; figure~\ref{geometrys} shows the view of the Geant4 geometry. The Geant4 Radioactive Decay Module was used for simulating decays, after checking carefully the energy conservation in the decay of all the considered isotopes. The low energy models based on Livermore data libraries were considered for the physical processes of $\alpha$, $\beta$ and $\gamma$ emissions. Uniformly distributed bulk contamination in the components was assumed, except if otherwise stated. For each simulated event, defined considering an energy integration time of 1 $\mu$s, the energy deposited at each detector by different types of particles has been recorded separately in order to build afterwards the energy spectrum, filtering alpha deposits above 2.5~MeV (as it can be made in real data by pulse shape analysis) and correcting each component with the corresponding Relative Scintillation Efficiency Factor\footnote{A constant value of 0.6 has been taken as relative efficiency factor for alpha particles in the building of the electron equivalent energy spectra. Energy from nuclear recoils is neglected.}. Production of scintillation at the NaI(Tl) crystals and the subsequent light collection have not been simulated here. Simulated spectra have been convoluted with gaussian functions to include the effect of the energy resolution, as characterized for the detectors at the different energy ranges \cite{anaiscompanion}. Energy spectra for different conditions have been constructed to allow direct comparison to data obtained from detectors.

Simulations have been scaled just by using the activities (or derived upper limits) given in tables \ref{extcon} and \ref{intcon}; for cosmogenic isotopes, the activity corresponding to different times of measurement is properly deduced from the production rates summarized in table~\ref{iso} or directly from the deduced activities for each detector in the case of $^{22}$Na and $^{3}$H (see sections~\ref{secna} and \ref{sech}).

\section{Background models and validation} \label{comparison}

\begin{figure*}
\centering
 \includegraphics[width=\textwidth]{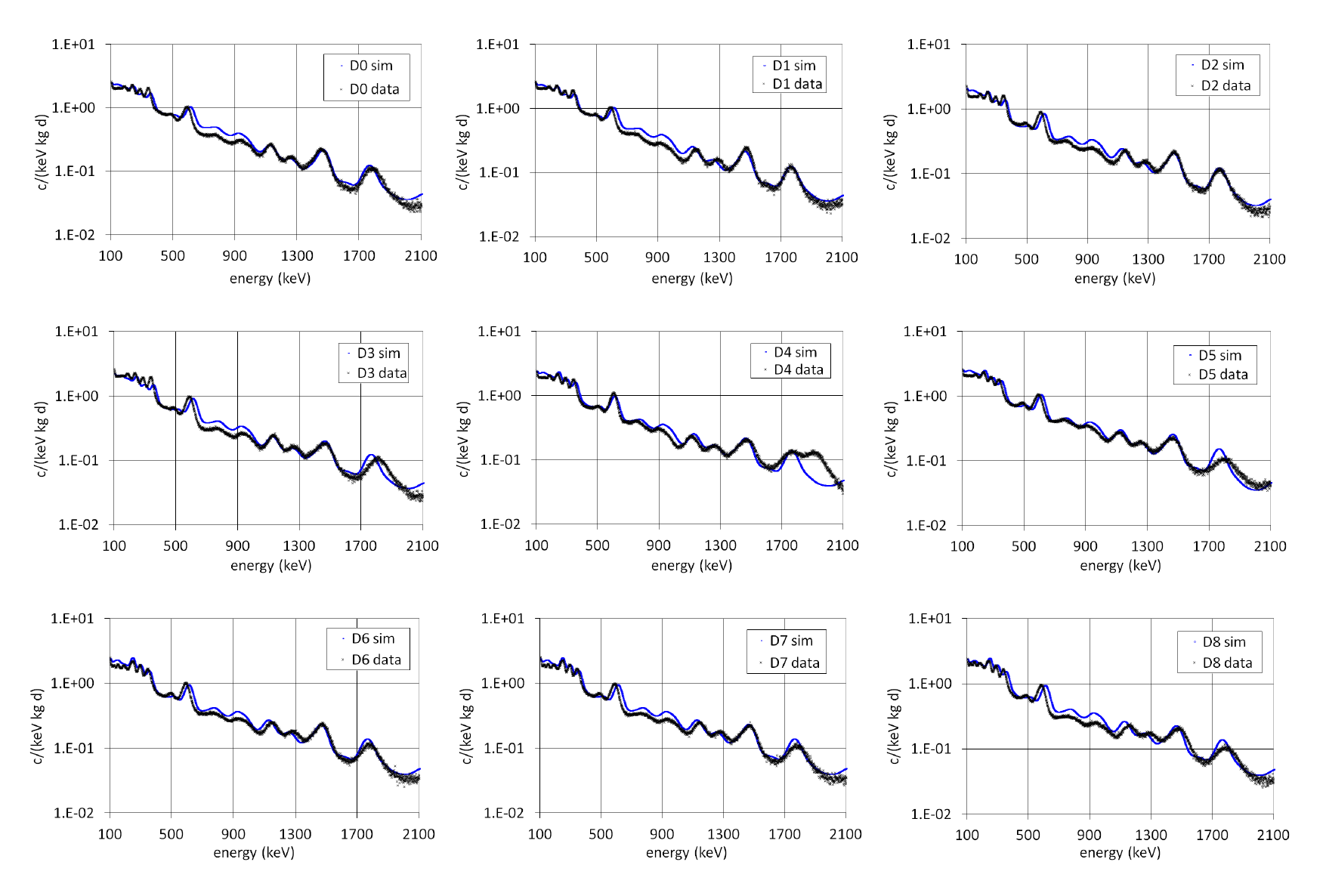}
 \caption{Comparison of the total energy spectra measured in the first year of data taking for each detector with the corresponding background model summing all the simulated contributions. Note that for detector D4 the PSA cannot perfectly work as high energy events (alpha particles and muons) are saturated and show equivalent energy below 2~MeV (see text).}
  \label{comparisonhet}
\end{figure*}

\begin{figure*}
\centering
 \includegraphics[width=\textwidth]{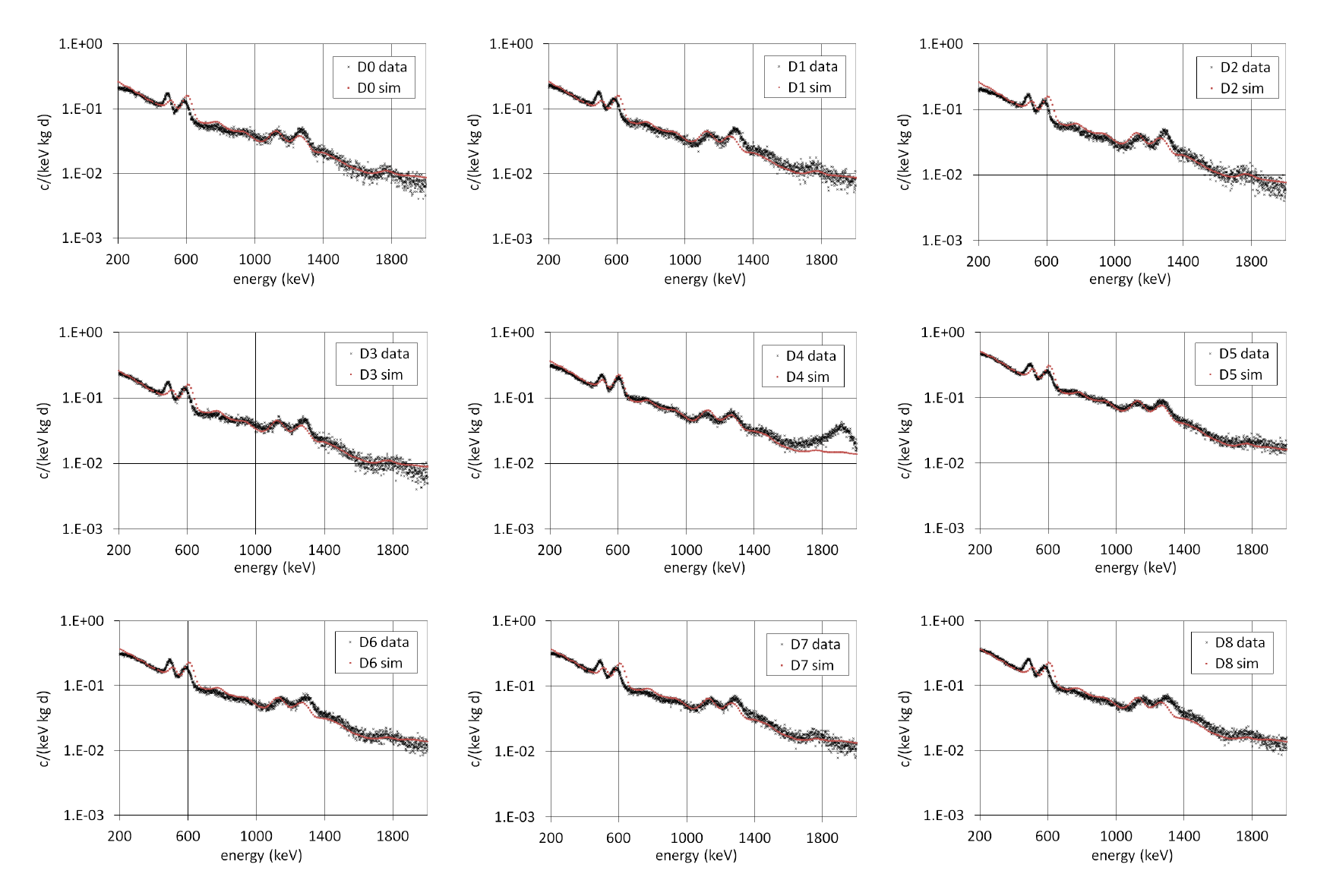}
 \caption{Comparison of the coincidence energy spectra measured in the first year of data taking for each detector with the corresponding background model summing all the simulated contributions, including the effect of Cerenkov emissions from the radioactivity of PMTs. Note that for detector D4 the PSA cannot perfectly work as high energy events (alpha particles and muons) are saturated and show equivalent energy below 2~MeV (see text).}
  \label{comparisonhec}
\end{figure*}

\begin{figure*}
\centering
 \includegraphics[width=\textwidth]{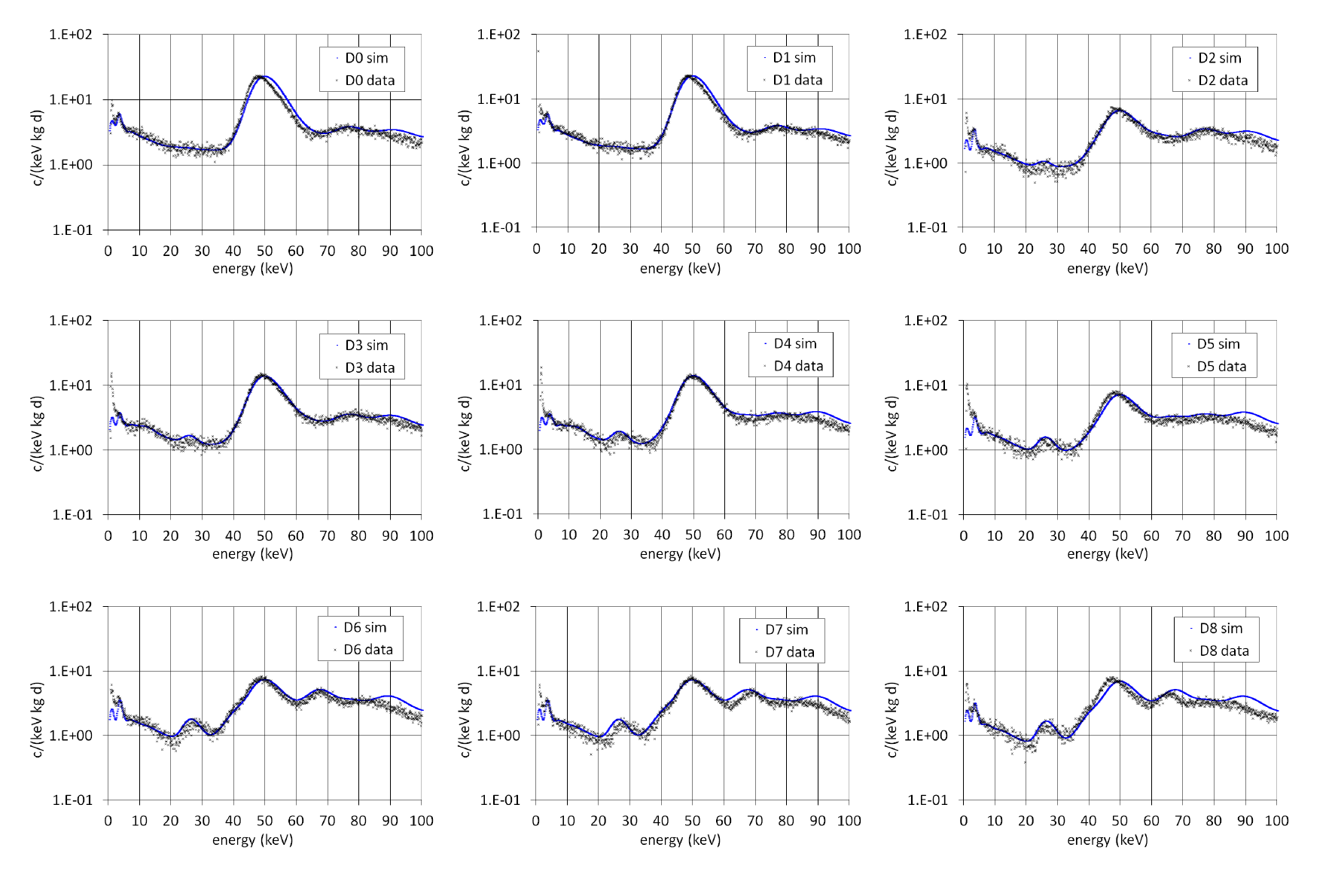}
 \caption{Comparison of the anticoincidence energy spectra in the low energy range up to 100~keV measured for each detector with the corresponding background model summing all the simulated contributions. The shown data correspond to the 10\% of unblinded data of the first year of data taking after filtering and efficiency correction.}
  \label{comparisonlea}
\end{figure*}

\begin{figure*}
\centering
 \includegraphics[width=\textwidth]{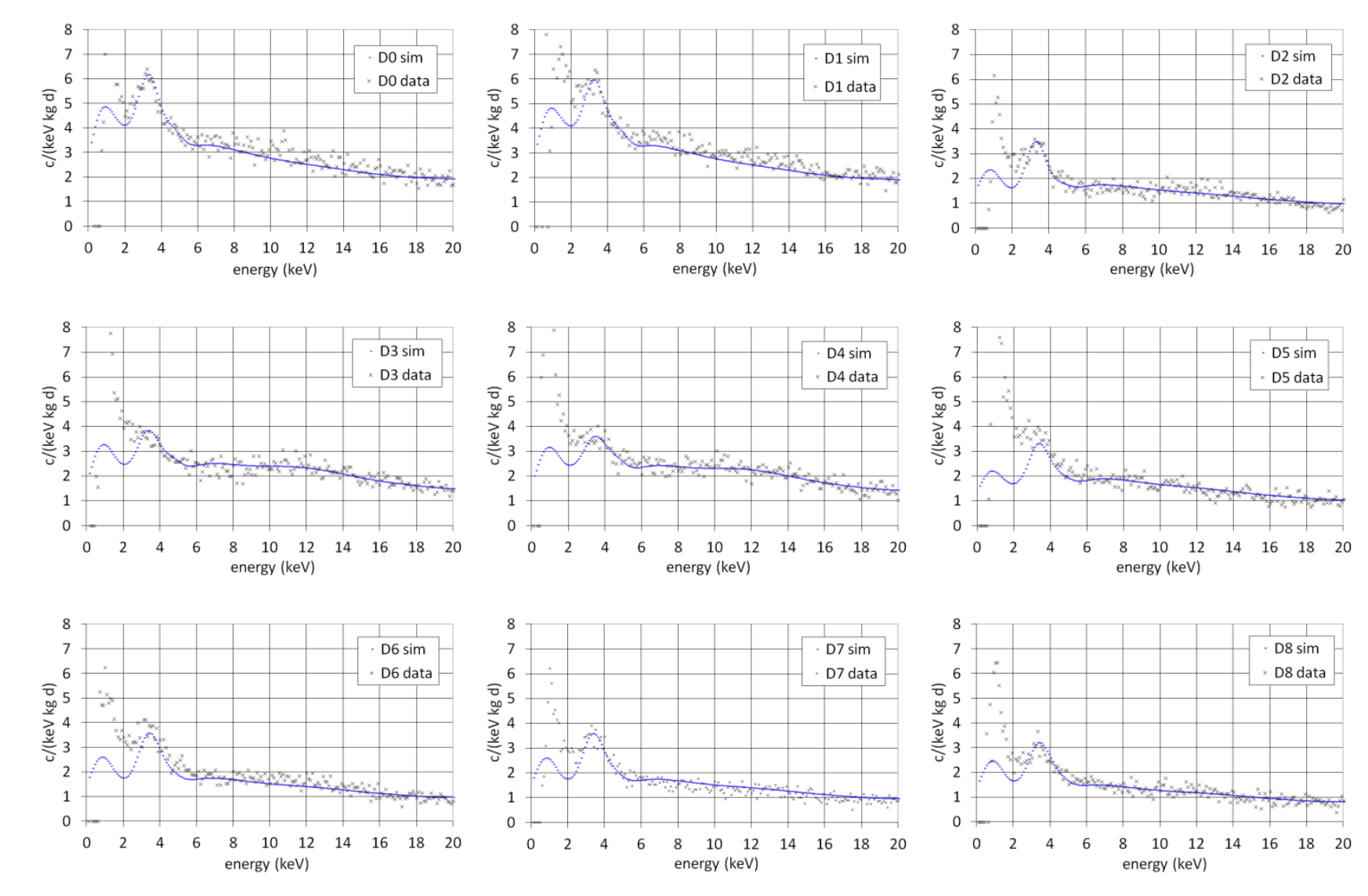}
 \caption{Comparison of the anticoincidence energy spectra in the region of interest measured for each detector with the corresponding background model summing all the simulated contributions. The shown data correspond to the 10\% of unblinded data of the first year of data taking after filtering and efficiency correction. These plots are just a zoom up to 20~keV of those displayed in figure~\ref{comparisonlea}.}
  \label{comparisonvlea}
\end{figure*}

The energy spectra measured in ANAIS-112 for each one of the nine detectors during the first year of data taking have been compared with the background model produced summing all the simulated contributions, in different conditions and energy ranges. Figures~\ref{comparisonhet} and \ref{comparisonhec} compare the results in the high energy region up to $\sim$2~MeV for total and coincidence spectra, respectively. Coincidence spectra include events having energy deposited at more than one NaI(Tl) detector. It must be noted that for detector D4 the PSA cannot work as high energy events are saturated, having an abnormal pulse shape and a pulse-area non-proportional to deposited energy, which implies a loss of "visible energy"; hence, alpha and muon events at very high energy contribute at or even below 2~MeV (see figures~\ref{comparisonhet} and \ref{comparisonhec} for D4). Table~\ref{ratesHE} presents the integrated measured rates for each detector in ANAIS-112 in this high energy region, together with the corresponding simulation and the quantified deviations; the statistical uncertainty of simulated rates is at the level or below 0.1\%. The comparison for the low energy region is specifically shown in figure~\ref{comparisonlea}, for anticoincidence spectra using the 10\% of unblinded data of the first year of data taking; a zoom of the region of interest up to 20~keV is presented in figure~\ref{comparisonvlea}. Table~\ref{ratesLE} presents the measured rates for each detector in ANAIS-112 in the region from 1 to 6~keV, corresponding to the 10\% of unblinded data, in comparison with the corresponding simulation and deviations; the statistical uncertainty of simulated rates for these low energy windows is at the level or below 1\%. The efficiency-corrected background level in the energy range from 1 to 6~keV is 3.58$\pm$0.02~keV$^{-1}$ kg$^{-1}$ d$^{-1}$. The overall agreement between data and simulation is quite satisfactory in general.

Since upper limits on radionuclide activity have been used for several components, the background could be overestimated in the models in some energy regions; in particular, a clear overestimation of the simulation around 92~keV has been suppressed by reducing the $^{238}$U upper limit for the copper vessel and quartz windows to that of $^{226}$Ra\footnote{Upper limits from gamma spectroscopy for the activity of isotopes at the upper part of the $^{238}$U chain are typically much larger than those at the lower part starting on $^{226}$Ra (as it can be seen in table~\ref{extcon}) because of the very low intensity of the gamma emissions at that chain segment.}. The inclusion of cosmogenic activation in the crystal (even if its decay is at different levels in different crystals) is essential to reproduce many features of the registered data; $^{22}$Na has a very important contribution in coincidence spectra. For the very low energy region, as discussed in sections~\ref{seccdsn} and \ref{sech}, $^{113}$Sn, $^{109}$Cd and $^{3}$H emissions allow to explain the peaks and the continuum level measured.



\begin{table*}
\begin{center}
\caption{Measured total rates (from figure~\ref{comparisonhet}) and rates of coincidences (from figure~\ref{comparisonhec}) up to 2~MeV for each ANAIS-112 detector and on average, from the first year of data taking. The corresponding simulated rates and their  deviation from the measurement are also presented (see text).}
\begin{tabular}{@{}c|ccc|ccc@{}} \hline 
& & total rate, 0.1-2~MeV & & & coincidences, 0.2-2~MeV &  \\ \hline
Detector & Measurement & Simulation & Deviation & Measurement & Simulation & Deviation  \\
& (kg$^{-1}$ d$^{-1}$) & (kg$^{-1}$ d$^{-1}$) & (\%) & (kg$^{-1}$ d$^{-1}$) & (kg$^{-1}$ d$^{-1}$) & (\%) \\
 \noalign{\smallskip}\hline\noalign{\smallskip}
D0 &  992.6$\pm$0.5 & 1048.0 & 5.6 & 102.3$\pm$0.2 & 108.9 & 6.5 \\
D1 & 1000.4$\pm$0.5  & 1038.6 & 3.8 & 107.0$\pm$0.2 & 108.9 & 1.7 \\
D2 & 798.8$\pm$0.4 & 842.8 & 5.5 & 99.1$\pm$0.2 & 106.9 & 7.8 \\
D3 &  920.2$\pm$0.5 & 910.9 & -1.0 & 107.3$\pm$0.2 & 109.1 & 1.6 \\
D4 & 956.9$\pm$0.5 & 1012.5 & 5.8 & 156.7$\pm$0.2 & 158.4 & 1.1	\\
D5 &   1010.2$\pm$0.5 & 1082.8 & 7.2 & 215.9$\pm$0.2 & 216.9 & 0.5\\
D6 & 929.1$\pm$0.5 & 989.7 & 6.5 & 154.5$\pm$0.2 & 158.3 & 2.4\\
D7 &   909.7$\pm$0.5 & 990.8 & 8.9 & 152.2$\pm$0.2 & 159.0 & 4.5\\
D8 &  904.8$\pm$0.5 & 976.8 & 8.0 & 159.3$\pm$0.2 & 158.9 & -0.3 \\ \hline
ANAIS-112 &  935.8$\pm$0.1 & 988.1 & 5.6 & 139.4$\pm$0.1 & 142.8 & 2.9\\
\noalign{\smallskip}\hline\noalign{\smallskip}
\end{tabular} \label{ratesHE}
\end{center}
\end{table*}

\begin{table*}
\begin{center}
\caption{Measured rates (after correction for efficiency of the filtering procedure) in the regions from 1 to 2~keV and from 2 to 6~keV for each ANAIS-112 detector and on average, from the 10\% of unblinded data of the first year of data taking. The corresponding simulated rates and their deviation from the measurement are also presented (see text).}
\begin{tabular}{@{}c|ccc|ccc@{}} \hline 
& & 1 to 2~keV & & & 2 to 6~keV &  \\ \hline
Detector & Measurement & Simulation & Deviation & Measurement & Simulation & Deviation  \\
& (keV$^{-1}$ kg$^{-1}$ d$^{-1}$) & (keV$^{-1}$ kg$^{-1}$ d$^{-1}$) & (\%) & (keV$^{-1}$ kg$^{-1}$ d$^{-1}$) & (keV$^{-1}$ kg$^{-1}$ d$^{-1}$) & (\%) \\
 \noalign{\smallskip}\hline\noalign{\smallskip}
D0 &  6.62$\pm$0.18 & 4.37 & -34 & 4.58$\pm$0.05 & 4.53 & -1.0 \\
D1 & 6.55$\pm$0.20  & 4.36 & -33 & 4.66$\pm$0.05 & 4.46 & -4.4 \\
D2 & 3.62$\pm$0.14 & 1.84 & -49 & 2.44$\pm$0.04 & 2.27 & -7.0 \\
D3 &  6.40$\pm$0.17 & 2.77 & -57 & 3.16$\pm$0.04 & 2.97 & -6.2 \\
D4 & 5.54$\pm$0.16 & 2.73 & -51 & 3.12$\pm$0.04 & 2.88 & -7.6	\\
D5 &   5.84$\pm$0.16 & 1.84 & -68 & 2.96$\pm$0.04 & 2.34 & -20.9\\
D6 & 4.16$\pm$0.16 & 2.04 & -51 & 2.90$\pm$0.04 & 2.42 & -16.3\\
D7 &   3.78$\pm$0.13 & 2.03 & -46 & 2.61$\pm$0.04 & 2.42 & -7.4\\
D8 &  3.74$\pm$0.13 & 1.94 & -48 & 2.29$\pm$0.04 & 2.18 & -5.1 \\ \hline
ANAIS-112 &  5.14$\pm$0.05 & 2.66 & -48 & 3.19$\pm$0.01 & 2.94 & -7.9\\
\noalign{\smallskip}\hline\noalign{\smallskip}
\end{tabular} \label{ratesLE}
\end{center}
\end{table*}


As shown in figure~\ref{comparisonlea}, $^{210}$Pb emissions at $\sim$50~keV are well reproduced assuming the activity as deduced from the alpha rate. To fully reproduce the shape of the very low energy region of the measured spectra it has been also necessary to consider a part of the quantified $^{210}$Pb activity in the NaI(Tl) crystals not in the bulk but on the surface, which could be related with $^{222}$Rn deposited once the crystals had been grown. As shown in \cite{anaisbkg}, $^{210}$Pb emissions from the crystal bulk or from the surface considering different depths up to 100~$\mu$m produce very different features below the peak at $\sim$50~keV, according to Monte Carlo simulation. For ANAIS-112 detectors, the fraction and depth of the $^{210}$Pb surface emission have been fixed for each crystal as needed to reproduce the registered low energy data, and vary from 25 to 75\% and 30 to 100~$\mu$m.  Additionally, a $^{210}$Pb activity in the teflon diffuser surrounding the crystals has been considered for some of the modules (D3 and D4); a value of 3~mBq/detector has been chosen to reproduce the structure
observed in the low energy background at $\sim$12~keV in some detectors. It is worth pointing out that changes in the production protocols were implemented during the manufacture of the different detectors by Alpha Spectra in the process of reducing the $^{210}$Pb activity in the crystals. Possible correlations of the features of the $^{210}$Pb surface emissions at each detector with the activity value or with the peak structure observed for the alpha emissions due to the $^{210}$Po (at the decay sequence of $^{210}$Pb) not showing the pure structure expected from a crystal bulk $^{210}$Pb contamination\footnote{The range of a 5-MeV alpha particle in NaI is 29~$\mu$m, following NIST data \cite{nist}. Then, for the $^{210}$Po alpha particle, fully absorbed for bulk emissions, a continuum at the left side of the peak is expected to appear only for very thin surface layers, which is not observed in ANAIS modules.} have been investigated \cite{anaisbkg} but no firm conclusion has been drawn and the issue is still under study.

\begin{figure*}
\centering
 \includegraphics[width=0.6\textwidth]{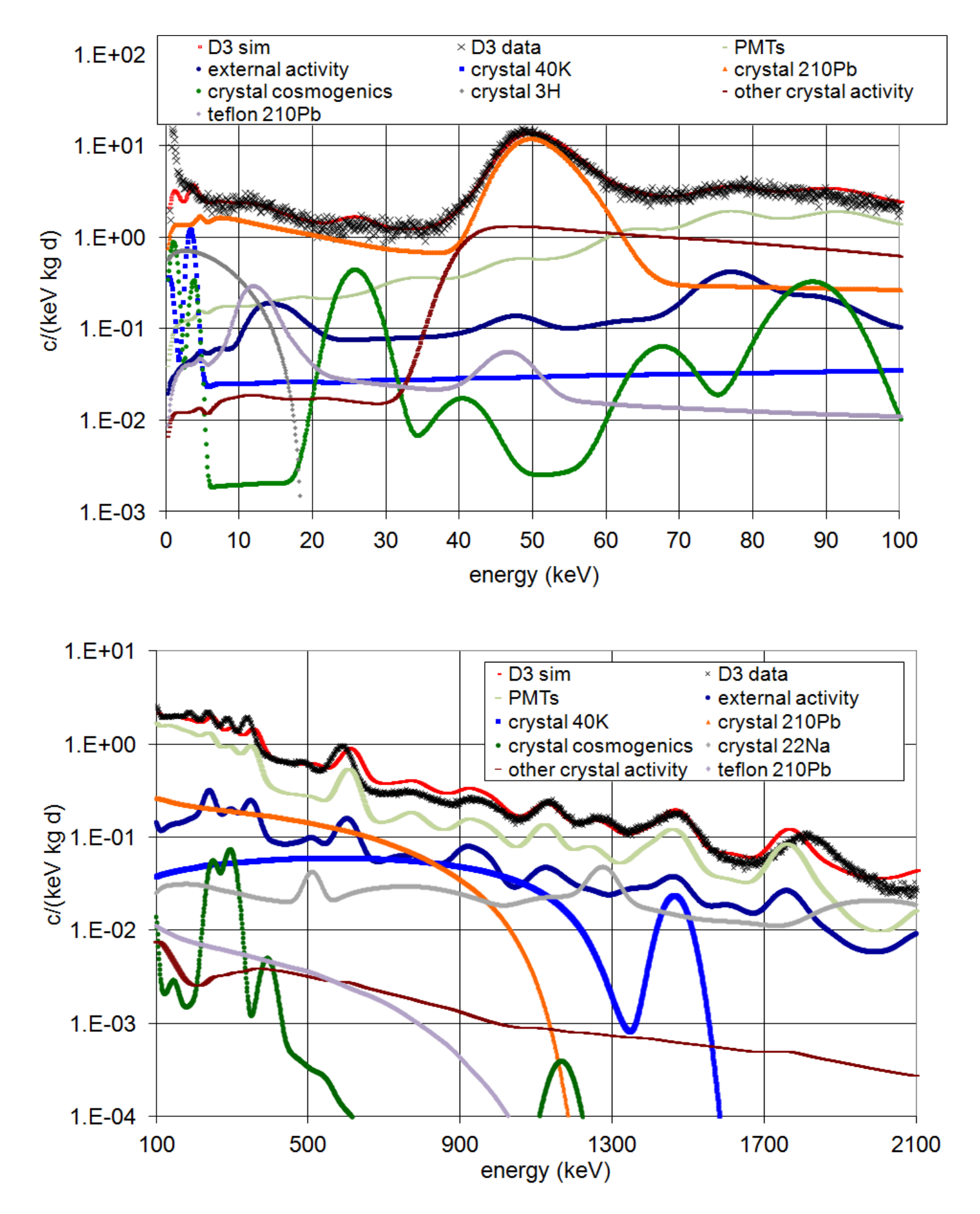}
 \caption{Energy spectra expected from individual background sources, together with the sum of all of them and the measured background. Results are presented for D3 detector, for anticoincidence spectra in the low energy range as in figure~\ref{comparisonlea} (top) and for total spectra in the high energy region as in figure~\ref{comparisonhet} (bottom). The measurement and simulations correspond to the first year of data taking. The shown low energy data correspond to the 10\% of unblinded data for that period.}
  \label{contributions}
\end{figure*}

Figure~\ref{contributions} presents, for one of the ANAIS-112 detectors as an example, the energy spectra expected from different background sources both in the low and high energies. While photomultiplier tubes' activity is the most relevant one in the high energy range, contributions from crystals, $^{210}$Pb and $^{3}$H continua and $^{40}$K, $^{22}$Na and $^{109}$Cd peaks, are the most significant for the region of interest at very low energies.

\begin{figure*}
\centering
 \includegraphics[width=1\textwidth]{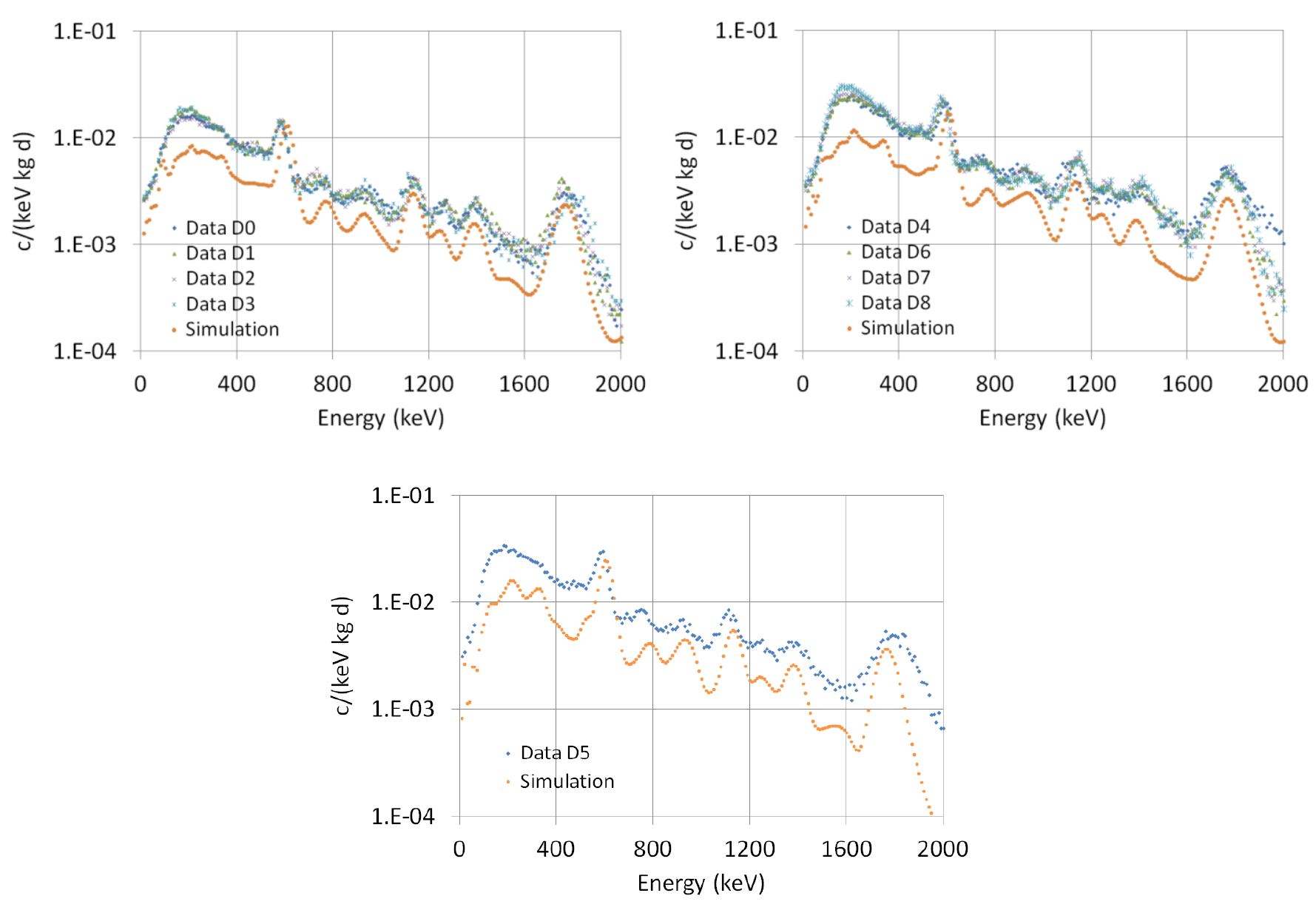}
 \caption{Energy spectra in the detector having an event compatible with NaI scintillation for the selected coincidence events with a very low energy anomalous event (rejected by the filtering procedure) in another one detector coming supposedly from PMT radioactive contaminations. Measured data taken in ANAIS-112 during the first year of data taking are compared with the corresponding simulation (see text). In the plots detectors are grouped according to their location in the 3$\times$3 matrix (see figure~\ref{anais112setup}, right).}
  \label{pmts}
\end{figure*}

The background simulation framework developed for ANAIS-112 has allowed to investigate some particular event populations identified in the analysis of the data. Effective filtering protocols have been developed to reject non-scintillation events, which limit the energy threshold, based on multiparametric cuts to properly select events with pulse shapes from NaI(Tl) scintillation \cite{anaiscompanion}. A part of the rejected events are thought to be generated by Cerenkov emission produced by the radioactive contaminations present in the photomultipliers (and quantified as summarized in table~\ref{extcon}). This effect should give in addition coincidences between two detectors when the gamma emissions produced by the PMT contaminations reach a different detector; these events have been selected in the ANAIS-112 data looking for coincidences having a very low energy anomalous event (rejected by the filtering procedure) in one detector and any energy deposit in another one. As the energy spectra observed for these selected events was compatible with the one expected from PMT emissions, dominated by $^{226}$Ra, a simulation was undertaken to check this hypothesis. The photocathode, the borosilicate glass and the copper enclosure of ANAIS-112 PMTs were included in the simulation; energy absorbed in the PMTs was recorded and the threshold for Cerenkov light production by electrons was applied to generate a coincidence. The borosilicate glass has refractive index n$=$1.517; the Cerenkov threshold calculated for n$=$1.5 is 685~keV. Figure~\ref{pmts} compares the measured energy spectra (in the detector having an event compatible with NaI scintillation) for the selected coincidence events, together with the corresponding simulation for all the radioactive isotopes quantified in the photomultipliers (supposed to be uniformly distributed in the PMTs); plots are shown for all the detectors, grouped according to their location in the 3$\times$3 matrix (see figure~\ref{anais112setup}, right). The mean values of activity shown in table~\ref{extcon} have been used for scaling the simulated spectra. The simulation reproduces the shape and roughly the counting rates at the main peaks of the measured spectra, which could support the hypothesis of having anomalous events which are rejected coming from PMT radioactivity; the discrepancy for the continuum levels could be due to a partial non-validity of some of the assumptions (Cerenkov threshold, uniform radioactive contaminations, ideal detection in simulation, \dots). This contribution from the PMTs radioactivity has indeed been considered as another background source in the simulation of coincidence spectra presented in figure~\ref{comparisonhec}, which allows to better reproduce the peak at 1764.5~keV. Moreover, an estimate of the rate in anticoincidence spectra from Cerenkov events from all the quantified PMT activity has been made from the simulation giving $\sim$300~kg$^{-1}$ d$^{-1}$; this value agrees with the measured rate of rejected events attributable to this effect.

The background models developed agree very well with the data above 2~keV but below this energy there is some discrepancy. From table~\ref{ratesLE}, it can be seen that the model does not explain 48\% of the measured rate in the region from 1 to 2~keV, but the deviation is of only 7.9\% from 2 to 6~keV. In the high energy region up to 2000~keV, the model agrees at 5.6\% (2.9\%) for total (coincidence) rates, giving indeed a slight overestimation. The unexplained events below 2~keV could be attributed to non-bulk scintillation events which have not been rejected or to background sources not included in the described model. Nuclear recoils following $^{210}$Po decays from surface crystal or teflon contamination could be relevant, but the unknown Relative Efficiency Factor of the nuclei makes difficult to quantify their contribution. On the other hand, $^{210}$Pb decay from external surface contaminations would produce much different structures in the low energy spectrum, which are not observed in present data. Results from the blank module in operation since the beginning of the second year of data taking will help to check some of these hypotheses.

\section{Conclusions} \label{conclusions}


The background of all the nine detectors used in the ANAIS-112 set-up has been thoroughly studied using the available data of the dark matter run, together with previous data taken at the smaller  set-ups with the first detectors built. Background models for each detector have been developed, based on Monte Carlo simulations and the measured activity in external components and in crystals, including cosmogenic products, quantified in dedicated, independent measurements based on different analysis techniques. The background models of ANAIS-112 detectors provide a good overall description of measured data at all energy ranges above 2~keV and at different analysis conditions (coincidence or anticoincidence).

After the full data analysis presented in \cite{anaiscompanion}, the overall background level from 1 to 6~keV measured for ANAIS-112 (from the 10\% of unblinded data of the first year of data taking) is 3.58$\pm$0.02~keV$^{-1}$ kg$^{-1}$ d$^{-1}$. The background in this region of interest for dark matter searches is totally dominated by crystal emissions. In particular, $^{210}$Pb and $^{3}$H continua and $^{40}$K, $^{22}$Na and $^{109}$Cd peaks are the most significant contributions. $^{40}$K and $^{22}$Na peaks are partially reduced thanks to anticoincidence operation in the 3$\times$3 modules matrix. According to the background model developed, $^{210}$Pb, $^{3}$H and $^{40}$K account for 32.5\%, 26.5\% and 12.0\% of the measured rate in ANAIS-112 from 1 to 6~keV respectively. During the production of the nine ANAIS-112 modules, improved purification and surface machining protocols were progressively implemented thanks to the interaction with Alpha Spectra and the availability of the first background data from operation at the Canfranc Underground Laboratory of some detectors; this led to a reduction of crystal activity, very significant for $^{210}$Pb and moderate for $^{40}$K. The directly quantified activities roughly explain the observed background, but the inclusion of some other components significantly improves the agreement between simulated and real data:
\begin{itemize}
\item partial crystal surface $^{210}$Pb contamination is needed to reproduce the low energy spectra of all modules, even though in a different amount and in different depth profile.
\item $^{3}$H in the crystal, which cannot be disentangled from other backgrounds and independently quantified, is also required to explain the measured continuum at very low energy. Indeed, the fully-absorbed emissions from $^{3}$H could become the main contribution in the region of interest; therefore, a shielding against cosmogenic activation was procured for the production of the last ANAIS NaI(Tl) crystals.
\end{itemize}

The comparison of selected coincidence data with the corresponding simulations finding good agreement has allowed to confirm relevant issues in the analysis of the low energy events to be used to look for annual modulation effects, like the ability of properly registering events at 1~keV or the origin of rejected events due to Cerenkov light emission from PMT radioactive contaminations. More work is underway in order to better understand the unexplained events below 2~keV, which could be related either with non-bulk scintillation events which have not been rejected by our filtering procedure, or some background which has not been taken into consideration in our model; in particular, we are considering nuclear recoils following alpha decays in surface crystal contaminations and loss of efficiency in light collection on crystal surface layers.

Following the sensitivity prospects presented in \cite{ivan}, ANAIS-112 could detect the annual modulation observed by DAMA/LIBRA or, otherwise, exclude the 3$\sigma$ region singled out by DAMA/LIBRA in five years. The good understanding of the background of ANAIS-112 detectors is important not only to make reliable background estimates in the blinded region, but also to have under control possible background-related systematics in the dark matter annual modulation analysis.


\begin{acknowledgements}
Professor J.A. Villar passed away in August, 2017. Deeply in sorrow, we all thank his dedicated work and kindness. This work has been financially supported by the Spanish Ministerio de Econom\'{i}a y Competitividad and the European Regional Development Fund (MINECO-FEDER) under grants No. FPA2014-55986-P and FPA2017-83133-P, the Consolider-Ingenio 2010 Programme under grants MultiDark CSD2009-00064 and CPAN CSD2007-00042 and the Gobierno de Arag\'{o}n and the European Social Fund (Group in Nuclear and Astroparticle Physics and I. Coarasa predoctoral grant). We thank student Alvaro Mart\'{i}n Miram\'{o}n for his work on cosmogenic isotopes and Jos\'{e} Mar\'{i}a L\'{o}pez Guti\'{e}rrez from Centro Nacional de Aceleradores, Sevilla, Spain for the analysis of $^{129}$I content. We acknowledge also the technical support from LSC and GIFNA staff.
\end{acknowledgements}


\begin{thebibliography}{99}
\bibitem{bernabei13} R.~Bernabei et al., Final model independent result of DAMA/LIBRA-phase1, Eur. Phys. J. C {\bf 73}, 2648 (2013)
\bibitem{bernabei18} R.~Bernabei et al., First model independent results from DAMA/LIBRA-phase2, Nucl. Phys. At. Energy {\bf 19}, 307 (2018)
\bibitem{cdmsresults} R.~Agnese et al. (SuperCDMS Collaboration), Results from the Super Cryogenic Dark Matter Search (SuperCDMS) experiment at Soudan, Phys. Rev. Lett. {\bf 120}, 061802 (2018)
\bibitem{edelweissresults} L.~Hehn et al. (EDELWEISS Collaboration), Improved EDELWEISS-III sensitivity for low-mass WIMPs using a profile likelihood approach, Eur. Phys. J C {\bf 76}, 548 (2016)
\bibitem{cresstresults} G.~Angloher et al. (CRESST Collaboration), Results on light dark matter particles with a low-threshold CRESST-II detector, Eur. Phys. J C {\bf 76}, 25 (2016); F.~Petricca et al. (CRESST Collaboration), First results on low-mass dark matter from the CRESST-III experiment, arXiv:1711.07692 [astro-ph.CO]
\bibitem{xenonresults} E.~Aprile et al. (XENON Collaboration), Dark Matter Search Results from a One Tonne$\times$Year Exposure of XENON1T, Phys. Rev. Lett. {\bf 121}, 111302 (2018)
\bibitem{luxresults} D.S.~Akerib et al. (LUX Collaboration), Results from a Search for Dark Matter in the Complete LUX Exposure, Phys. Rev. Lett. {\bf 118}, 021303 (2017)
\bibitem{pandaxresults} X. Cui et al. (PandaX-II Collaboration), Dark Matter Results From 54-Ton-Day Exposure of PandaX-II Experiment, Phys. Rev. Lett. {\bf 119}, 181302 (2017)
\bibitem{darksideresults} P.~Agnes et al. (The DarkSide Collaboration), Constraints on Sub-GeV Dark-Matter Electron Scattering from the DarkSide-50 Experiment, Phys. Rev. Lett. {\bf 121}, 111303 (2018); Low-Mass Dark Matter Search with the DarkSide-50 Experiment, Phys. Rev. Lett. {\bf 121}, 081307 (2018)
\bibitem{picoresults} C.~Amole et al., Dark Matter Search Results from the PICO-60 C$_{3}$F$_{8}$ Bubble Chamber, Phys. Rev. Lett. {\bf 118}, 251301 (2017)
\bibitem{damicresults}A.~Aguilar-Arevalo et al. (DAMIC Collaboration),Search for low-mass WIMPs in a 0.6 kg day exposure of the DAMIC experiment at SNOLAB, Phys. Rev. D {\bf 94}, (2016) 082006
\bibitem{newsgresults} Q.~Arnaud et al. (NEWS-G Collaboration), First results from the NEWS-G direct dark matter search experiment at the LSM, Astropart. Phys. {\bf 97}, 54 (2018)
\bibitem{kang} S.~Kang et al., DAMA/LIBRA-phase2 in WIMP effective models, JCAP {\bf 07}, 016 (2018)
\bibitem{freese} K.~Freese et al., Annual Modulation of Dark Matter: A Review, Rev. Mod. Phys. {\bf 85}, 1561 (2013)
\bibitem{xenonmodulation} E.~Aprile et al. (XENON Collaboration),  XENON Search for Event Rate Modulation in XENON100 Electronic Recoil Data, Phys. Rev. Lett. {\bf 115}, 091302 (2015)
\bibitem{xmassmodulation} K.~Abe, et al. (XMASS Collaboration), Direct dark matter search by annual modulation in XMASS-I, Phys. Lett. B {\bf 759}, 272 (2016)
\bibitem{luxmodulation} D.S.~Akerib et al., A search for annual and diurnal rate modulations in the LUX experiment, Phys. Rev. D {\bf 98}, 062005 (2018)
\bibitem{cogentmodulation} C.E.~Aalseth et al. (GoGeNT Collaboration), CoGeNT: A Search for Low-Mass Dark Matter using p-type Point Contact Germanium Detectors, Phys. Rev. D {\bf 88}, 012002 (2013)
\bibitem{fushimi99} K.~Fushimi et al., Limits on the annual modulation of WIMPs nucleus scattering with large-volume NaI(Tl) scintillators, Astropart. Phys. {\bf 12}, 185 (1999)
\bibitem{sarsa97} M.L.~Sarsa et al., Results of a search for annual modulation of WIMP signals, Phys. Rev. D {\bf 56}, 1856 (1997)
\bibitem{bernabei99} R.~Bernabei et al., Performances of the 100 kg NaI (Tl) set-up of the DAMA experiment at Gran Sasso, Riv. Nuovo Cim. A {\bf 112}, 545 (1999)
\bibitem{gerbier99} G.~Gerbier et al., Pulse shape discrimination with NaI(Tl) and results from a WIMP search at the Laboratoire Souterrain de Modane, Astropart. Phys. {\bf 11}, 287 (1999)
\bibitem{naiad} G.J.~Alner et al., Limits on WIMP cross-sections from the NAIAD experiment at the Boulby Underground Laboratory, Phys. Lett. B {\bf 616}, 17 (2005)
\bibitem{anais17} J.~Amar\'e et al., The ANAIS-112 experiment at the Canfranc Underground Laboratory, arXiv:1710.03837v1 [physics.ins-det]
\bibitem{cosine} G.~Adhikari et al., Initial performance of the COSINE-100 experiment, Eur. Phys. J. C {\bf 78}, 107 (2018) 
\bibitem{cosinenature} The COSINE-100 Collaboration, An experiment to search for dark-matter interactions using sodium iodide detectors, Nature {\bf 564}, 83 (2018); Author Correction: An experiment to search for dark-matter interactions using sodium iodide detectors, Nature {\bf 566}, E2 (2019)
\bibitem{kims} P.~Adhikari et al., Understanding internal backgrounds in NaI(Tl) crystals toward a 200 kg array for the KIMS-NaI experiment,  Eur. Phys. J. C {\bf 76}, 185 (2016) 
\bibitem{dmice} J.~Cherwinka et al., First Search for a Dark Matter Annual Modulation Signal with NaI(Tl) in the Southern Hemisphere by DM-Ice17, Phys. Rev. D {\bf 95}, 032006 (2017) 
\bibitem{sabre} C.~Tomei et al. (SABRE Collaboration), SABRE: Dark matter annual modulation detection in the northern and southern hemispheres, Nucl. Instrum. Meth. A {\bf 845}, 418 (2016)
\bibitem{cosinus} G.~Angloher et al., Results from the first cryogenic NaI detector for the COSINUS project, JINST {\bf 12}, P11007 (2017)
\bibitem{xenon100bkg} E.~Aprile et al., Study of the electromagnetic background in the XENON100 experiment, Phys. Rev. D {\bf 83}, 082001 (2011)
\bibitem{edelweissbkg} E.~Armengaud et al., Background studies for the EDELWEISS dark matter experiment, Astropart. Phys. {\bf 47}, 1 (2013)
\bibitem{gerdabkg} M.~Agostini et al., The background in the neutrinoless double beta decay experiment GERDA, Eur. Phys. J. C {\bf 74}, 2764 (2014)
\bibitem{luxbkg} D.S.~Akerib et al., Radiogenic and Muon-Induced Backgrounds in the LUX Dark Matter Detector, Astropart. Phys. {\bf 62}, 33 (2015)
\bibitem{exobkg} J.B.~Albert et al., Investigation of radioactivity-induced backgrounds in EXO-200, Phys. Rev. C {\bf 92}, 015503 (2015)
\bibitem{cuorebkg} C.~Alduino et al. (CUORE Collaboration), The projected background for the CUORE experiment, Eur. Phys. J. C {\bf 77}, 543 (2017)
\bibitem{cosinebkg} G.~Adhikari et al., Understanding NaI(Tl) crystal background for dark matter Searches, Eur. Phys. J. C {\bf 77}, 437 (2017) P.~Adhikari et al., Background model for the NaI(Tl) crystals in COSINE-100, Eur. Phys. J. C {\bf 78}, 490 (2018)
\bibitem{anaisjcap} J.~Amar\'e et al., Cosmogenic radionuclide production in NaI(Tl) crystals, JCAP {\bf 02}, 046 (2015)
\bibitem{anaisbkg} J.~Amar\'e et al, Assessment of backgrounds of the ANAIS experiment for dark matter direct detection, Eur. Phys. J. C {\bf 76}, 429 (2016)
\bibitem{anaisap} S.~Cebri\'an et al., Background model for a NaI (Tl) detector devoted to dark matter searches, Astropart. Phys. {\bf 37}, 60 (2012)
\bibitem{anaisepjc} C.~Cuesta et al., Bulk NaI(Tl) scintillation low energy events selection with the ANAIS--0 module, Eur. Phys. J. C {\bf 74}, 3150 (2014)
\bibitem{anaisnima} J.~Amar\'e et al., Preliminary results of ANAIS--25, Nucl.Instrum. Meth. A {\bf 742}, 197 (2014) 
\bibitem{anaispprocedia} J. Amar\'e et al., From ANAIS-25 towards ANAIS-250, Physics Procedia {\bf 61}, 157 (2015)
\bibitem{anaisyield} M.A.~Oliv\'an et al., Light yield determination in large sodium iodide detectors applied in the search for dark matter, Astrop. Phys. {\bf 93}, 86 (2017)
\bibitem{MAthesis} M.A.~Oliv\'an, Design, scale-up and characterization of the data acquisition system for the ANAIS dark matter experiment, PhD thesis, Universidad de Zaragoza, (2016), arXiv:1601.07312 [physics.ins-det]
\bibitem{anaiscompanion} J.~Amar\'e et al., Performance of ANAIS-112 experiment after the first year of data taking, Eur. Phys. J. C {\bf 79}, 228 (2019)
\bibitem{Clarathesis} C.~Cuesta, ANAIS--0: Feasibility study for a 250~kg NaI(Tl) dark matter search experiment at the Canfranc Underground Laboratory, PhD thesis, Universidad de Zaragoza (2013)
\bibitem{Carmona:2004qk} J.M. Carmona et~al., Neutron background at the Canfranc underground laboratory and its contribution to the IGEX-DM dark matter experiment, Astropart. Phys. \textbf{21}, 523 (2004)
\bibitem{Jordan:2013} D.~Jordan et~al., Measurement of the neutron background at the Canfranc Underground Laboratory LSC, Astropart. Phys. \textbf{42}, 1 (2013)
\bibitem{bettini2012} A. Bettini, The Canfranc Underground Laboratory, Eur. Phys. J. Plus {\bf 127}, 112 (2012)
\bibitem{anaisijmpa} C.~Cuesta et al., Analysis of the $^{40}$K contamination in NaI(Tl) crystals from different providers, Int. J. Mod. Phys. A {\bf 29}, 1443010 (2014)
\bibitem{bernabei2008} R.~Bernabei et al., The DAMA/LIBRA apparatus, Nucl. Instrum. Meth. A {\bf 592}, 297 (2008)
\bibitem{dec1} http://www.nucleide.org/DDEP.htm
\bibitem{dec2} WWW Table of Radioactive Isotopes, http://nucleardata.nuclear.lu.se/toi/
\bibitem{dmicethesis} W.C.~Pettus, Cosmogenic Activation in NaI Detectors for Dark Matter Searches, PhD thesis, University of Wisconsin-Madison (2015)
\bibitem{ijmpaNaI} P.~Villar et al., Study of the cosmogenic activation in NaI(Tl) crystals within the ANAIS experiment, Int. J. Mod. Phys. A {\bf 33}, 1843006 (2018)
\bibitem{sabrebkg} M.~Antonello et al., Monte Carlo simulation of the SABRE PoP background, Astropart. Phys. \textbf{106}, 1 (2019)
\bibitem{tritio} J.~Amar\'e et al., Cosmogenic production of tritium in dark matter detectors, Astropart. Phys. {\bf 97}, 96 (2018)
\bibitem{pvillar} P.~Villar, Background evaluation of the ANAIS dark matter experiment in different configurations: towards a final design, PhD Dissertation, Universidad de Zaragoza, 2016,
https://zaguan.unizar.es/record/58561/files/TESIS- 2017-007.pdf
\bibitem{tendl} A.J.~Koning, D.~Rochman, Modern nuclear data evaluation with the TALYS code system, Nuclear Data Sheets {\bf 113}, 2841 (2012)
\bibitem{head} Y.A.~Korovin, High energy activation data library (HEAD-2009), Nucl. Instrum. Meth. A {\bf 624}, 20 (2010)
\bibitem{gordon} M.S.~Gordon, Measurement of the flux and energy spectrum of cosmic-ray in- duced neutrons on the ground, IEEE Trans. Nucl. Sci. {\bf  51}, 3427 (2004). Erratum: M.S.~Gordon et al., IEEE Transactions on Nuclear Science {\bf 52}, 2703 (2005)
\bibitem{geant4} S.~Agostinelli et al., Geant4: a simulation toolkit, Nucl. Instrum. Meth. A {\bf 506}, 250 (2003)
\bibitem{nist} National Institute of Standards and Technology, http://physics.nist.gov/PhysRefData/Star/Text/ASTAR.html
\bibitem{ivan} I. Coarasa et al., ANAIS-112 sensitivity in the search for dark matter annual modulation, Eur. Phys. J. C {\bf 79}, 233 (2019)

\end{thebibliography}
\end{document}